\documentclass[a4paper, 12pt]{article}
\usepackage{bm}
\usepackage{amsfonts}
\usepackage{amsmath}
\usepackage{amssymb}
\usepackage{amsthm}
\usepackage{authblk}
\usepackage{color}
\usepackage{listings}
\usepackage{xcolor}
\usepackage[utf8]{inputenc}
\usepackage[T1]{fontenc}
\counterwithin{equation}{section}
\usepackage[margin=2.0cm]{geometry}
\usepackage{graphicx}
\usepackage{float}
\usepackage{subcaption}
\usepackage{tabularx}
\usepackage{url}
\usepackage{appendix}

\begin{document}

\title{A Deep Reinforcement Learning Trader without Offline Training}
\author[1]{Boian Lazov\thanks{blazov\_fte@uacg.bg}}
\affil[1]{Department of Mathematics, University of Architecture, Civil Engineering and Geodesy, 1164 Sofia, Bulgaria}

\maketitle

\begin{abstract}
\noindent In this paper we pursue the question of a fully online trading algorithm (i.e. one that does not need offline training on previously gathered data). For this task we use Double Deep $Q$-learning in the episodic setting with Fast Learning Networks approximating the expected reward $Q$. Additionally, we define the possible terminal states of an episode in such a way as to introduce a mechanism to conserve some of the money in the trading pool when market conditions are seen as unfavourable. Some of these money are taken as profit and some are reused at a later time according to certain criteria. After describing the algorithm, we test it using the 1-minute-tick data for Cardano's price on Binance. We see that the agent performs better than trading with randomly chosen actions on each timestep. And it does so when tested on the whole dataset as well as on different subsets, capturing different market trends.
\end{abstract}

\section{Introduction}

In recent years algorithmic trading on financial markets is increasingly replacing humans \cite{treleaven13}. One can find numerous estimates for the market share of automated traders with some sources giving over $73\%$ for US equity trading \cite{treleaven13}, while others citing as high as $92\%$ for forex trading.\footnote{See for example \url{https://www.quantifiedstrategies.com/what-percentage-of-trading-is-algorithmic/}.\\ Such sources, however, do not seem that reliable, since data is generally not openly available. Nevertheless, there are many (paid) reports that give a general idea of the scope of automated trading:\\ \url{https://www.grandviewresearch.com/industry-analysis/algorithmic-trading-market-report},\\ \url{https://www.mordorintelligence.com/industry-reports/algorithmic-trading-market}, \\ \url{https://www.alliedmarketresearch.com/algorithmic-trading-market-A08567}.}. Regardless of the actual figures, intelligent automation is increasingly used in our world and promises to be applicable in some very complex domains, where analytic solutions are either not known or very hard to obtain.\\
\indent There are many possible approaches to developing a trading algorithm, but recently one direction of research has been receiving much attention, namely machine learning based approaches. In particular, Reinforcement Learning (RL) has been a really promising way to solve some very difficult problems in other areas (like learning to play various games like Go \cite{silver16} and StarCraft II \cite{vinyals19} at the expert level) and is now being adapted to make decisions and execute trades in the trading setting. This area of research is very active and fairly new (see for example \cite{millea21}).\\
\indent As promising as it is, RL suffers from one big problem, namely the generalisation one (as does all of machine learning in fact). More specifically, once the agent (or neural network) is trained on a set of data and good performance is achieved, it is generally hard to translate this training to a new dataset and keep the performance. Furthermore, the training set usually needs to be very large and the agent needs to replay it many times. This is obviously not a good situation for a trading algorithm, since the market is considered a stochastic system and as such it changes rapidly and continuously. If we hope to be able to predict its movement, it should mostly be short term. For this the trader should be able to adapt quickly to current information.\\
\indent There are many proposed ways to try to deal with said problem of generalisation (both in supervised learning and RL), but one that seems both promising and simple is the idea to learn only the output weights of a neural network. It is implemented in partiular in two algorithms -- Extreme Learning Machine (ELM) \cite{huang04,ding14} and Fast Learning Network (FLM) \cite{li14}, and one can borrow the structure of the FLN to use as an approximator to the $Q$-function. As we will see later this will be a success and we will obtain a RL agent that performs better than random even when the market's overall trend is downward.\\
\indent This paper is organised as follows: in section \ref{alg} we will describe the algorithm in detail, namely all of the components of the RL (section \ref{algdq}) as well as the neural network (section \ref{algfln}); in section \ref{imptest} we will discuss briefly how to implement the algorithm and then we will test it on historical market data against an algorithm, that takes random actions on each timestep; finally, we will end with some concluding remarks (section \ref{conc}).

\section{The algorithm}\label{alg}

Our goal is to build a simple trading algorithm, which uses a pool of money to trade for some asset, by observing the state of the market. More precisely our agent needs to collect some information about the market, then open a position (ideally, executing a trade). Then it needs to wait for more information to calculate whether the trade was a good one and the process repeats.\\
\indent We will build our trader in the framework of RL. More precisely we will use a double $Q$-learning algorithm with approximation \cite{sutton18}. It is generally thought that combining $Q$-learning with approximation should be avoided, because of instabilities, but there are examples of successfully using such algorithms \cite{hasselt18,hasselt16}. To approximate the $Q$-functions we will use the structure of a FLN \cite{li14}. On top of that we will also propose a savings mechanism to deal with ``bearish'' markets. The idea is to take money out of the trading pool, so that part of it is never used again and can be taken as profit and another part is returned to the trading pool when the market conditions seem favourable. We will now go through all the specific components one by one.

\subsection{Double Q-learning}\label{algdq}

\subsubsection{State}

We will use a standard double Q-learning algorithm. It will be episodic with a continuous state space. As is usual, we will add an index $t$ to variables to denote the current time step and $t+1$ for the next time step. First we will see how to construct the state of the environment. As we will be using approximaiton of the $Q$-function by a neural network the state will be defined by a feature vector.\\
\indent The algorithm is initialised with 3 pools of money, denoted by $mon$, $sav$ and $res$. $mon$ refers to the current pool with which to trade. $sav$ denotes an amount of money, that are saved and never again used. This gives a convenient way to use the profit, without disturbing the operation of the trader and also safeguards somewhat against big losses. $res$ refers to a pool of money, that are stored for later use when some conditions are met. Finally, there are also the assets that the agent will buy, denoted by $cns$. $mon$ and $cns$ will be included in the calculation of the state.\\
\indent Next, we need a variable, which will be used to determine when to move money to $sav$ and $res$. We denote this by $mlim$. We will describe this in more detail later, but briefly $sav$ and $res$ will be increased (and $mon$ decreased), when the value of $mon$ becomes greater than $mlim$.\\
\indent To calculate the state, our trader also needs information for the market. It consists of one price, recorded at the beginning of the episode, denoted by $ipr$, and $5$ consecutive prices, denoted by $pr_1$, $pr_2$, $pr_3$, $pr_4$ and $pr_5$, recorded at some intervals.\\
\indent The next thing that is needed to calculate the state is the trading volume. More precisely, the trader records the volumes from the intervals immediately preceding the ones with recorded prices. These $5$ volumes are then used to calculate a few averages, that will be included in the feature vector. These are a simple moving average of the previous $100$ values of the volume, denoted by $av$ as well as the average volume of the current state's data points, denoted by $cav$. The last recorded volume (associated with $pr_5$) is also used in the feature vector and it is denoted by $vol_5$.\\
\indent Finally, we also include in the feature vector the Relative Strength Index (RSI), calculated with the recorded prices (15 prices, as is standard), as well as the relative movements of the price and relative movements of the movements and so on, i.e.
\begin{align}
nmd^{(1)}_i&=\frac{pr_{i+1}-pr_i}{pr_i},\ \ \ i\in\{1,2,3,4\},\\
nmd^{(2)}_k&=\frac{nmd^{(1)}_{k+1}-nmd^{(1)}_k}{nmd^{(1)}_k},\ \ \ k\in\{1,2,3\},\\
nmd^{(3)}_l&=\frac{nmd^{(2)}_{l+1}-nmd^{(2)}_l}{nmd^{(2)}_l},\ \ \ l\in\{1,2\},\\
nmd^{(4)}&=\frac{nmd^{(3)}_2-nmd^{(3)}_1}{nmd^{(3)}_1}.
\end{align}

\indent With all of the above the feature vector has the following form:
\begin{align}
feat= &\left(1,pr_1,...,pr_5,ipr,\frac{pr_5-ipr}{ipr}, mon, cns, cav, av,\frac{cav-av}{av},\frac{vol_5-av}{av},\right.\label{feature}\\
&\left.\frac{vol_5-cav}{cav},rsi,nmd^{(1)}_1,...,nmd^{(1)}_4,nmd^{(2)}_1,...,nmd^{(2)}_3,...,nmd^{(4)},mlim \right)^{T}.\notag
\end{align}

\subsubsection{Actions and rewards}

Next, we want to define the set of actions that the agent can take. They are simple -- buy, sell or hold. We denote the set of actions as $\{ a_1,a_2,...,a_{19} \}$. Actions $a_1$ to $a_9$ denote buying. Action $a_1$ means the agent buys coins for $10$ money, action $a_2$ -- for $20$ and so on. Actions $a_{10}$ to $a_{18}$ denote selling for a fixed amount of money -- again at increments of $10$. Finally, action $a_{19}$ denotes holding. Actions, of course, can fail due to insufficient funds and this will be reflected in the reward.\\
\indent This leads us to the next ingredient of the RL algorithm -- the reward signal. In order to calculate the reward we first need to keep a record of the total wealth $wth$ of the agent, associated with a given state,
\begin{align}
wth=mon+pr_5\ cns.\label{wth}
\end{align}

\noindent Thus after trading and recording the next $5$ prices, the wealth changes. Using this, we choose the reward to be polynomial in the change of the wealth, i.e.
\begin{align}
rew=(wth_{t+1}-wth_t)-\left(\frac{wth_{t+1}-wth_t}{2}\right)^2.
\end{align}

\noindent The above formula helps to discourage too risky actions (i.e. ones with changes in the wealth that are too big). Additionally, if the attempted action fails, the reward is instead
\begin{align}
rew=(wth_{t+1}-wth_t)-\left(\frac{wth_{t+1}-wth_t}{2}\right)^2-0.1.
\end{align}

\subsubsection{Terminal state}

As we mentioned earlier, we are using a savings mechanism. It ties in with the terminal state. We will consider three different terminal states. The first terminal state is the one in which
\begin{align}
mon_{t+1}>mlim,
\end{align}

\noindent where $mlim$ is a parameter that changes when encountering a terminal state. This means that the current money pool of the agent is greater than some threshold. Before beginning the new episode, the extra money $mdf=mon_{t+1}-mlim$ is distributed between three pools: $0.34$ goes to a savings pool $sav$, which is never again used to trade; $0.33$ -- to a reserves pool $res$, which might be used again later; and $0.33$ is left in the money pool (so that after this still $mon_{t+1}>mlim$). Then $mlim$ is increased to the current value of $mon$ plus $mdf$. The reward for going into this state is modified -- it is the usual plus the amount of money that was added to $sav$, i.e. $0.34\, mdf$.\\
\indent The second terminal state is determined by four conditions:
\begin{align}
&mon_{t+1}<mlim;\\
&wth_{t+1}<mlimn;\\
&Q(s_t,a_t)>0;\\
&rsi_{t+1}>70.
\end{align}

\noindent Here $mlimn$ is a hyperparameter, which sets the lowest possible value of $mlim$.  In general the meaning of RSI is open to interpretation \cite{wilder78, brown11}, but if all of the above conditions are met, we take this as an indication that the market conditions are favourable (while the agent is low on money), so half of the money in $res$ are redistributed to the money pool to be used for trading again. After this $mlim$ is again changed -- this time to $\max\{mlimn,mon_{t+1}+\frac{res}{2}\}$, and the new episode begins.\\
\indent The final terminal state is determined by the following:
\begin{align}
&mon_{t+1}<mlim;\\
&wth_{t+1}\ge mlimn;\\
&Q(s_t,a_t)<0;\\
&rsi_{t+1}<30.
\end{align}

\noindent Here the market is seen as unfavourable so the only thing to do is to change $mlim$ to $wth_{t+1}$, so that it will be easier to redistribute some of the money later.

\subsubsection{Policy and hyperparameters}\label{weights}

As is standard we use an $\varepsilon$-greedy policy. It picks actions based on the value of the average of the two $Q$-functions in a given state.  In general the choice of $\varepsilon$ is not a trivial task and the performance of the algorithm can vary greatly depending on this choice. There are many suggestions on how to successfully manage this balance of exploration and exploitation (using for example decay of $\varepsilon$ \cite{sutton18}, change point detection \cite{hartland07}, adaptation based on value differences \cite{tokic10}, etc.). What we want here is to be able to explore sufficiently when the market conditions change, which is very important for a fully online algorithm. In line with this we use a simple decay of $\varepsilon$, but mixed with a probabilistic reset to a larger value. More precisely, first initialise a counter $i_\varepsilon$ to $0$. Before each choice of an action $i_\varepsilon$ is either incremented by $1$ or with probability $prob_\varepsilon$ it is reset to $\left\lceil \frac{e^5-2}{5} \right\rceil$, if $i_\varepsilon\ge \left\lceil \frac{e^5-2}{5} \right\rceil$ (this resets $\varepsilon$ to about $0.2$). Afterwards, $\varepsilon$ is calculated according to the formula
\begin{align}
\varepsilon=\frac{1}{\ln\left(5i_\varepsilon+2\right)}.
\end{align}

\indent Next, we want to choose a learning rate $\alpha$. It is well-known that the learning rate in gradient descent methods greatly affects the performance of a neural network (or of the RL algorithm using it) \cite{smith17}. To avoid fixing the learning rate manually, we choose to use a cyclical one \cite{gulde20, gotmare19}. More precisely, a counter $i_\alpha$ is initialised to $0$. Then, before taking the previously chosen action $\alpha$ is calculated using the formula \cite{gotmare19}
\begin{align}
\alpha=\alpha_{min}+\frac{1}{2}\left( \alpha_{max}-\alpha_{min} \right)\left( 1+\cos\left( \frac{i_\alpha}{T_\alpha}\pi \right) \right),
\end{align}

\noindent after which $i_\alpha$ is incremented by $1$. This means that $\alpha$ varies between $\alpha_{max}$ and $\alpha_{min}$ with a period of $2T_\alpha$ steps.

\subsection{Fast learning network}\label{algfln}

\subsubsection{Preliminaries}

Now we need to describe the neural network, that will approximate the $Q$-function, namely FLN. FLNs use a parallel connection of two feedforward neural networks -- one has a single hidden layer, while the other has none \cite{li14}. The hidden layer weights are random and fixed and only the output weights are learned. If, in addition, we choose the output neurons' activation function to be the identity function and fix all the biases to zero, this effectively means that the approximating function is linear in the feature vector with additional fixed nonlinear terms from the hidden layer.\\
\indent More precisely, we can denote the input and output as $X$ and $Y$, respectively:
\begin{align}
X=\left( \begin{matrix} x_1 \\ x_2 \\ ... \\ x_n \end{matrix} \right),\ \ \ Y=\left( \begin{matrix} y_1 \\ y_2 \\ ... \\ y_m \end{matrix} \right),
\end{align}

\noindent where $n$ and $m$ are the respective sizes of the input and the output. Also, for shortness of notation, we denote the hidden layer output as
\begin{align}
g(Z)=\left( \begin{matrix} g(z_1) \\ g(z_2) \\ ... \\ g(z_r) \end{matrix} \right).
\end{align}

\noindent Here $g$ is the activation function of the hidden layer and $Z$ is the input to the hidden layer. $r$ is the hidden layer size.\\
\indent The weights are denoted by $W^{\mathrm{oi}}$ (input to output layers), $W^{\mathrm{hi}}$ (input to hidden layers) and $W^{\mathrm{oh}}$ (hidden to output layers):
\begin{align}
W^{\mathrm{oi}}=\left( \begin{matrix} w^{\mathrm{oi}}_{11} &  ... & w^{\mathrm{oi}}_{1n} \\ ... \\ w^{\mathrm{oi}}_{m1} & ... & w^{\mathrm{oi}}_{mn} \end{matrix} \right),\ \ \ W^{\mathrm{hi}}=\left( \begin{matrix} w^{\mathrm{hi}}_{11} & ... & w^{\mathrm{hi}}_{1n} \\ ... \\ w^{\mathrm{hi}}_{r1} & ... & w^{\mathrm{hi}}_{rn} \end{matrix} \right),\ \ \ W^{\mathrm{oh}}=\left( \begin{matrix} w^{\mathrm{oh}}_{11} & ... & w^{\mathrm{oh}}_{1r} \\ ... \\ w^{\mathrm{oh}}_{m1} & ... & w^{\mathrm{oh}}_{mr} \end{matrix} \right).
\end{align}

\indent Now the $k$-th component of the output vector is calculated by the following formula \cite{li14}:
\begin{align}
y_k=\sum_{s=1}^{n}w^{\mathrm{oi}}_{ks}x_s+\sum_{l=1}^{r}w^{\mathrm{oh}}_{kl}g\left( \sum_{t=1}^{n}w^{\mathrm{hi}}_{lt}x_{t} \right).
\end{align}

\noindent We can shorten the above to
\begin{align}
Y=W^{\mathrm{oi}}X+W^{\mathrm{oh}}g\left( W^{\mathrm{hi}}X \right).
\end{align}

\noindent The optimisation is then performed only with respect to $W^\mathrm{oi}$ and $W^\mathrm{oh}$.

\subsubsection{Weight renormalisation}

One common problem that we can encounter is that the weights in the neural network may diverge. This is especially true when the learning rate is large (but a large learning rate might help with adaptation). The above is a problem, since it is generally accepted that very large weights correlate with overfitting the training set and poor generalisation \cite{goodfellow16}. There are many ways to try to deal with this, but one simple method is to just renormalise the weight vector \cite{salimans16}. We do something similar with the output weights $W^\mathrm{oi}$ and $W^\mathrm{oh}$.\\
\indent More precisely, consider the output $y_k$. It is obtained by scalar multiplication of the weight vector
\begin{align}
\left(w^\mathrm{oi}_{k1},w^\mathrm{oi}_{k2},...,w^\mathrm{oi}_{kn},w^\mathrm{oh}_{k1},w^\mathrm{oh}_{k2},...,w^\mathrm{oh}_{kr}\right)^T\label{wv}
\end{align}

\noindent with the concatenation of the input $X$ and the hidden layer output $g(Z)$. The vector (\ref{wv}) itself is the concatenation of the $k$-th row of $W^\mathrm{oi}$ and the $k$-th row of $W^\mathrm{oh}$ and it is learned by stochastic gradient descent. A record of the maximal value of its norm $maxw$ is kept. If the weight vector is longer than $1$ after an update, it is rescaled by a factor of $\frac{1}{maxw}$. This keeps the weights from diverging and allows us to use large learning rates (the exact values of $\alpha_{min}$ and $\alpha_{max}$ are hyperparameters and will be specified later, but them being larger should help the agent adapt quickly).

\section{Implementation details and testing}\label{imptest}

\subsection{Observing, trading, hyperparameters}\label{obstr}

Before implementing the algorithm we need to consider a few points, namely how to record prices, how to trade, how exactly to structure the FLN and the values of the hyperparameters. The first question that needs to be answered is how often to record a price (and volume) for the feature vector. In principle the intervals can be of any length. One advantage of automated trading is that it can react quickly to the market. In line with this we want the intervals to be short, e.g. $1$ minute (more precisely the price is recorded in the beginning of the $1$-minute interval). However, a trade occurs right after observing $5$ prices, which in practice means that trades are performed every $5$ minutes. This means that, depending on the volatility of the market, consecutive trades might happen on similar (often the same) prices and the profit from this is very small. Possibly too small to compensate for the trading fee. To counter this the observed price passes through a filter before being recorded, such that the relative change between two prices is greater than $0.01$.\\
\indent The next question is how exactly to trade. In testing we just assume that the trade occurs at the last recorded price $pr_5$. To ensure this in practice one should use limit orders instead of market orders to avoid slippage. However, this poses the problem that the trade might not be executed at all (or at least not before new $5$ prices are recorded and it's time to trade again). To ensure that it has up-to-date information the trader should cancel the order before the next $5$ prices are recorded, e.g. after recording $pr_4$. Additionally, in such cases one can include the same negative reward as for trade failure due to insufficient funds to try to discourage orders that are later cancelled.\\
\indent Next, we need to describe the neural network in more detail. It's input is the feature vector (\ref{feature}), representing the state, while in its output we include one node for each action, i.e.
\begin{align}
X=feat,\ \ \ Y=\left( \begin{matrix} Q(s,a_1)\\ Q(s,a_2)\\...\\Q(s,a_{19}) \end{matrix} \right).
\end{align}

\noindent This means that there are $27$ input nodes and $19$ output nodes. The size of the hidden layer is a hyperparameter of the algorithm and it is fixed to $r=50$. Then the weight martices $W^\mathrm{oi}$, $W^\mathrm{hi}$ and $W^\mathrm{oh}$ are $19\times 27$, $50\times 27$ and $19\times 50$, respectively, while the weight vector (\ref{wv}) has $77$ components.\\
\indent From the above we see that there is a separate weight vector (\ref{wv}) for each action $a_k$. After choosing and taking the action $a_k$ only the respective weight vector should be updated. So the gradient of $Q(s,a_k)$ with respect to the weights is just the concatenation of $X$ and $g(Z)$ and it is the same for all actions.\\
\indent For the neuron activation function we choose to use the logistic function, i.e.
\begin{align}
g(z)=\frac{1}{1+e^{-z}},
\end{align}

\noindent and the feature vector (\ref{feature}) is scaled, so that its norm is $6$, before feeding it into the neural network.\\
\indent Finally, we need to fix the rest of the hyperparameters (in addition to the hidden layer size). For the discount factor we choose $\gamma=0.05$. The lowest possible value of $mlim$ is fixed to $mlimn=75$. While we have eliminated the need to choose $\varepsilon$, there is still a hyperparameter to fix and it is the probability for a reset of $\varepsilon$. We choose this to be $prob_\varepsilon=10^{-4}$. Likewise, we are not choosing the learning rate $\alpha$. Nevertheless, there are still hyperparameters to fix there also, namely $\alpha_{min}$, $\alpha_{max}$ and $T_\alpha$. We choose the following values: $\alpha_{min}=10^{-3}$, $\alpha_{max}=1$ and $T_\alpha=10^3$.

\subsection{Testing}

\begin{figure}[H]
\centering
\begin{subfigure}[t]{0.49\textwidth}
\includegraphics[width=\textwidth]{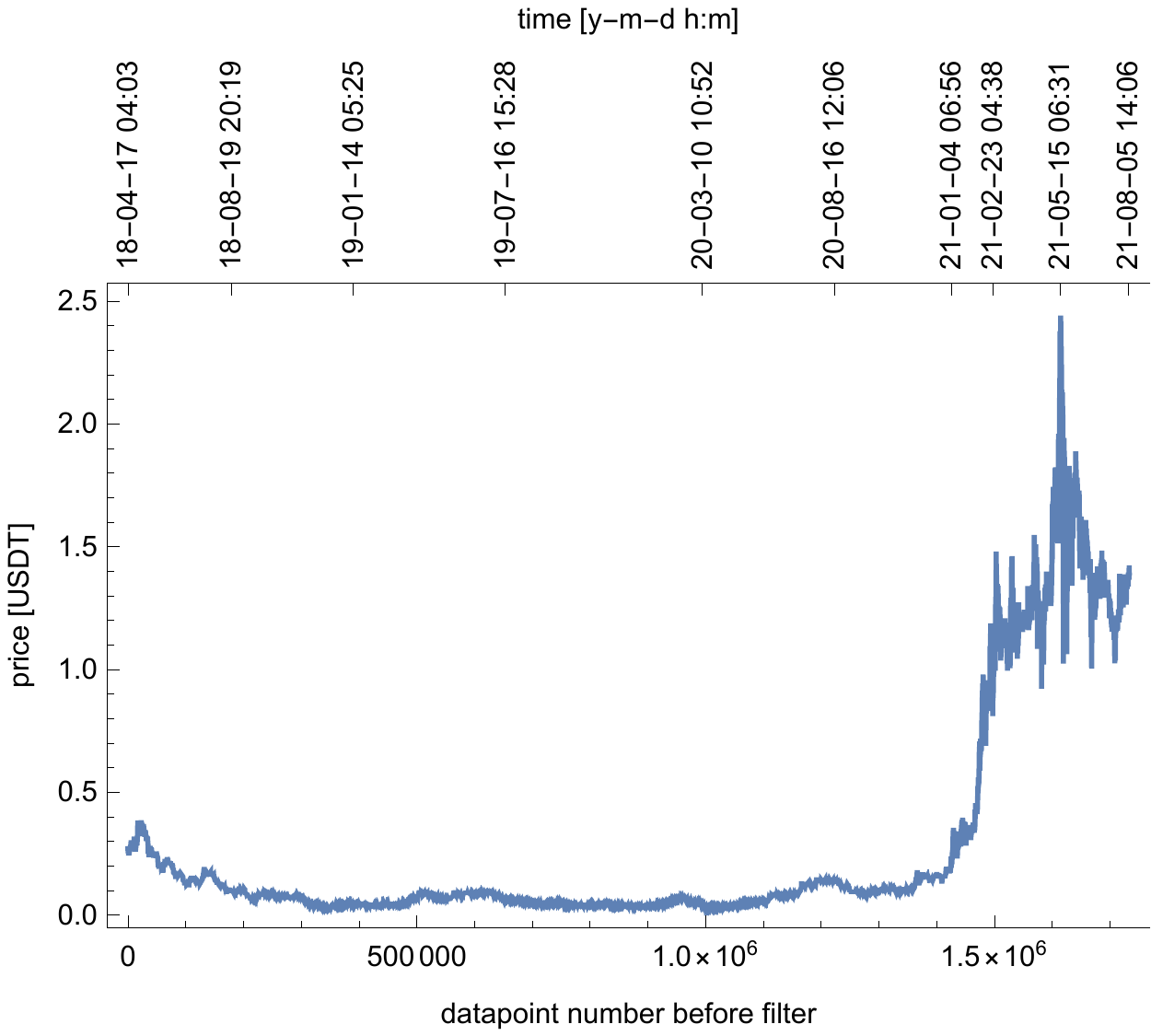}
\caption{The full dataset from $17.04.2018$ to $06.08.2021$.}
\label{prdatafull}
\end{subfigure}
\hspace{0.8 mm}
\begin{subfigure}[t]{0.49\textwidth}
\includegraphics[width=\textwidth]{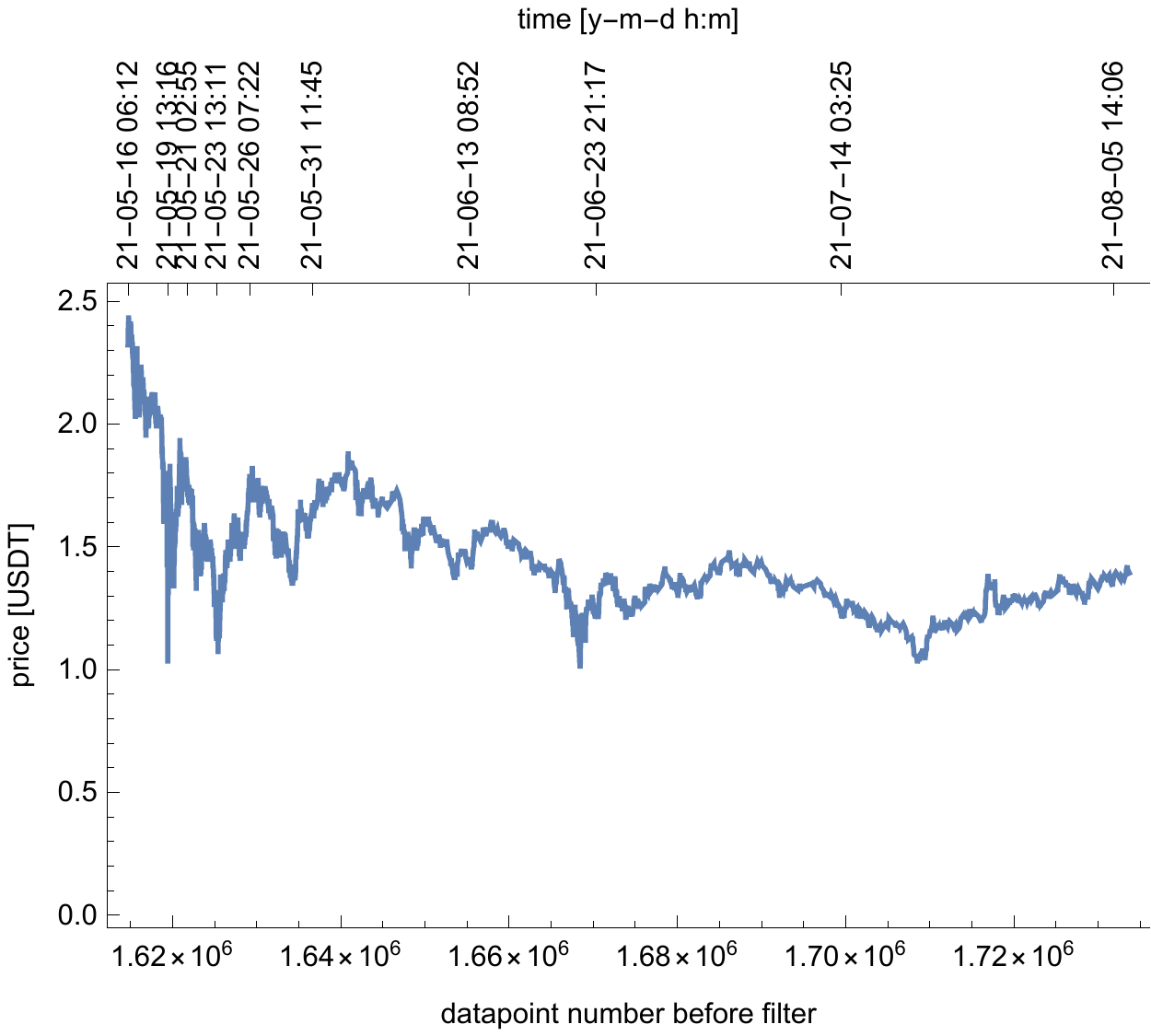}
\caption{A subset capturing a ``bearish'' market between $16.05.2021$ and $06.08.2021$.}
\label{prdatabear}
\end{subfigure}

\begin{subfigure}[t]{0.49\textwidth}
\includegraphics[width=\textwidth]{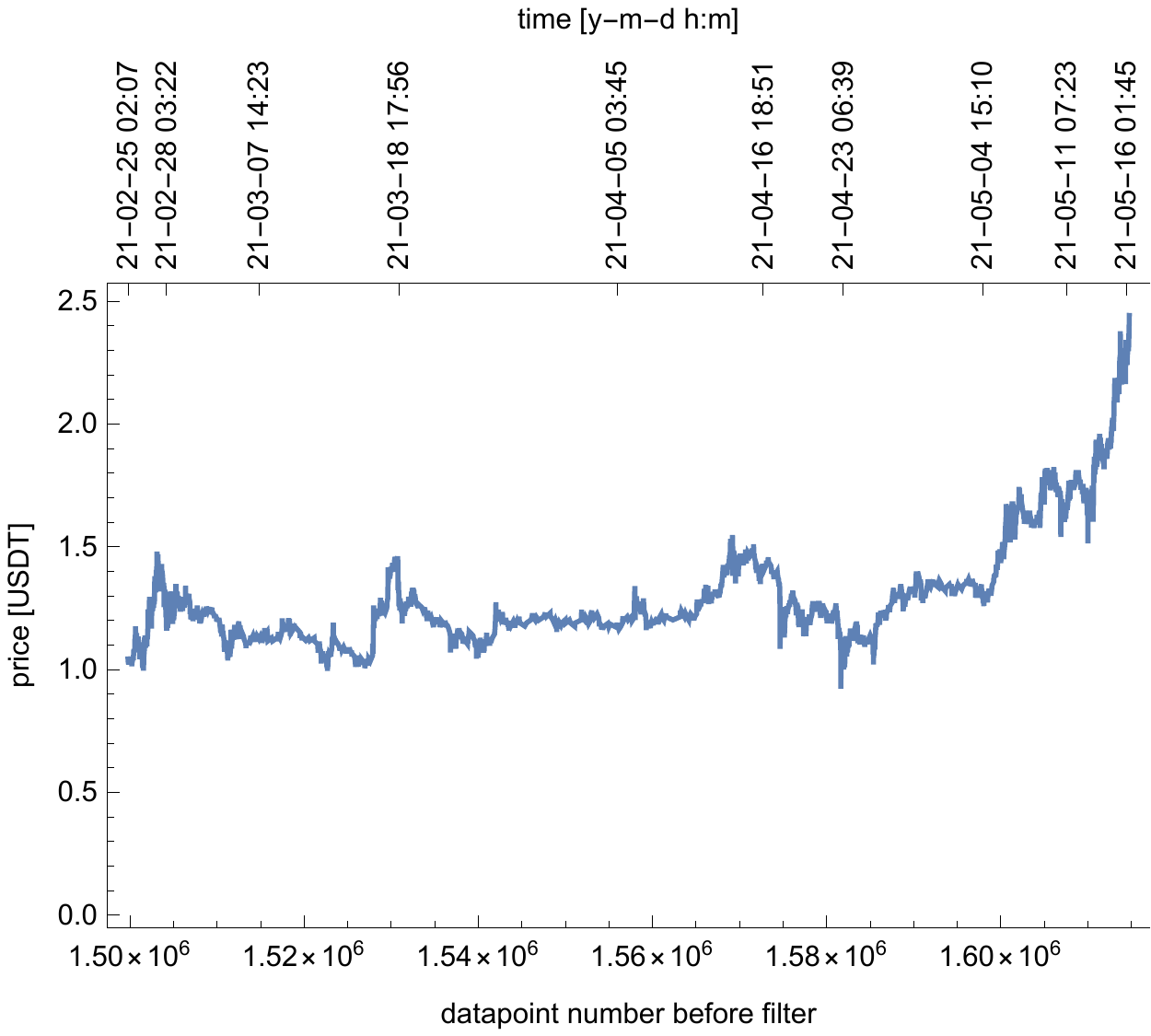}
\caption{A subset capturing a ``bullish'' market from $25.02.2021$ to $16.05.2021$.}
\label{prdatabull}
\end{subfigure}
\hspace{0.8 mm}
\begin{subfigure}[t]{0.49\textwidth}
\includegraphics[width=\textwidth]{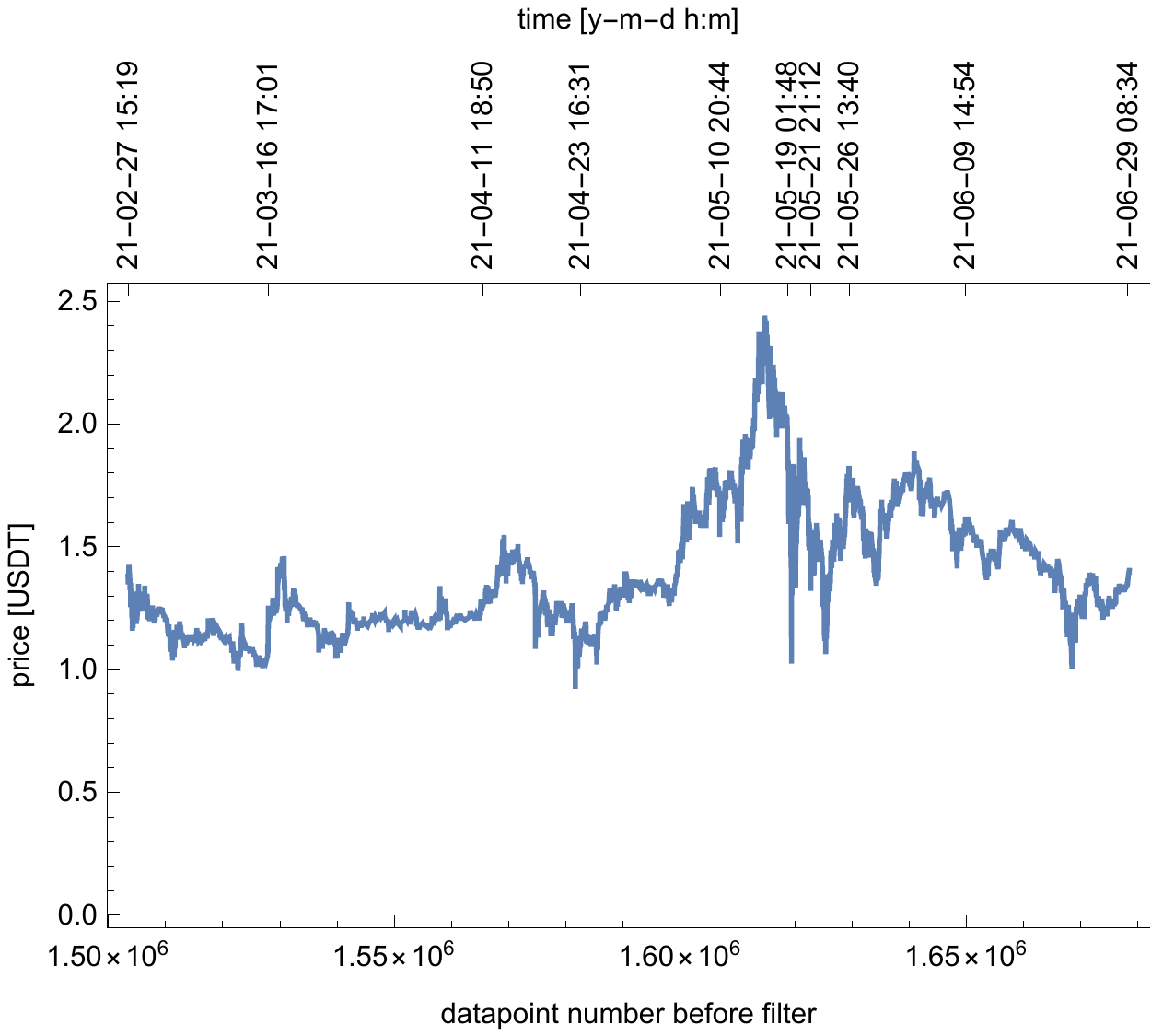}
\caption{A subset capturing a ``mixed'' market between $27.02.2021$ and $29.06.2021$.}
\label{prdatamix}
\end{subfigure}
\caption{Different subsets of the filtered dataset. The ticks along the upper frames are chosen to show the dates for 10 equally spaced (in terms of the datapoint number in the subset) points. A larger distance between two dates represents the lower volatility of the market (i.e. the smaller changes in price) in this period, which leads to more data points in the unfiltered data (and thus the greater distance) for the same amount of points in the filtered data.}\label{prdata}
\end{figure}

With the above considerations in mind here we present the results of testing the algorithm (the whole code for which is written in Mathematica and is included in appendix \ref{code}) on historical market data. Because we are using previously recorded prices, as already mentioned, there are a few things we can't account for, one of which is that in testing the order and the trade are the same, i.e. the order is always fully fulfilled at exactly the recorded price. Also, the precision is much higher when testing as we may use numbers with many digits. In real world applications one needs to round appropriately (e.g. when using part of $res$, when placing an order, etc.).\\
\indent In principle nothing stops us from usign the trader in any market, but our tests are performed on historical data for the ADA/USDT cryptocurrency pair on Binance from the pair listing on $17.04.2018$ to $06.08.2021$. We first pass the data through a filter as described in section \ref{obstr}. Then we use $4$ subsets of the filtered data. One is the whole dataset (figure \ref{prdatafull}), while the other three are attempts to capture different market conditions -- a ``bearish'' (figure \ref{prdatabear}), a ``bullish'' (figure \ref{prdatabull}) and a ``mixed'' (figure \ref{prdatamix}) market.

\begin{figure}[H]
\centering
\begin{subfigure}[t]{0.32\textwidth}
\includegraphics[width=\textwidth]{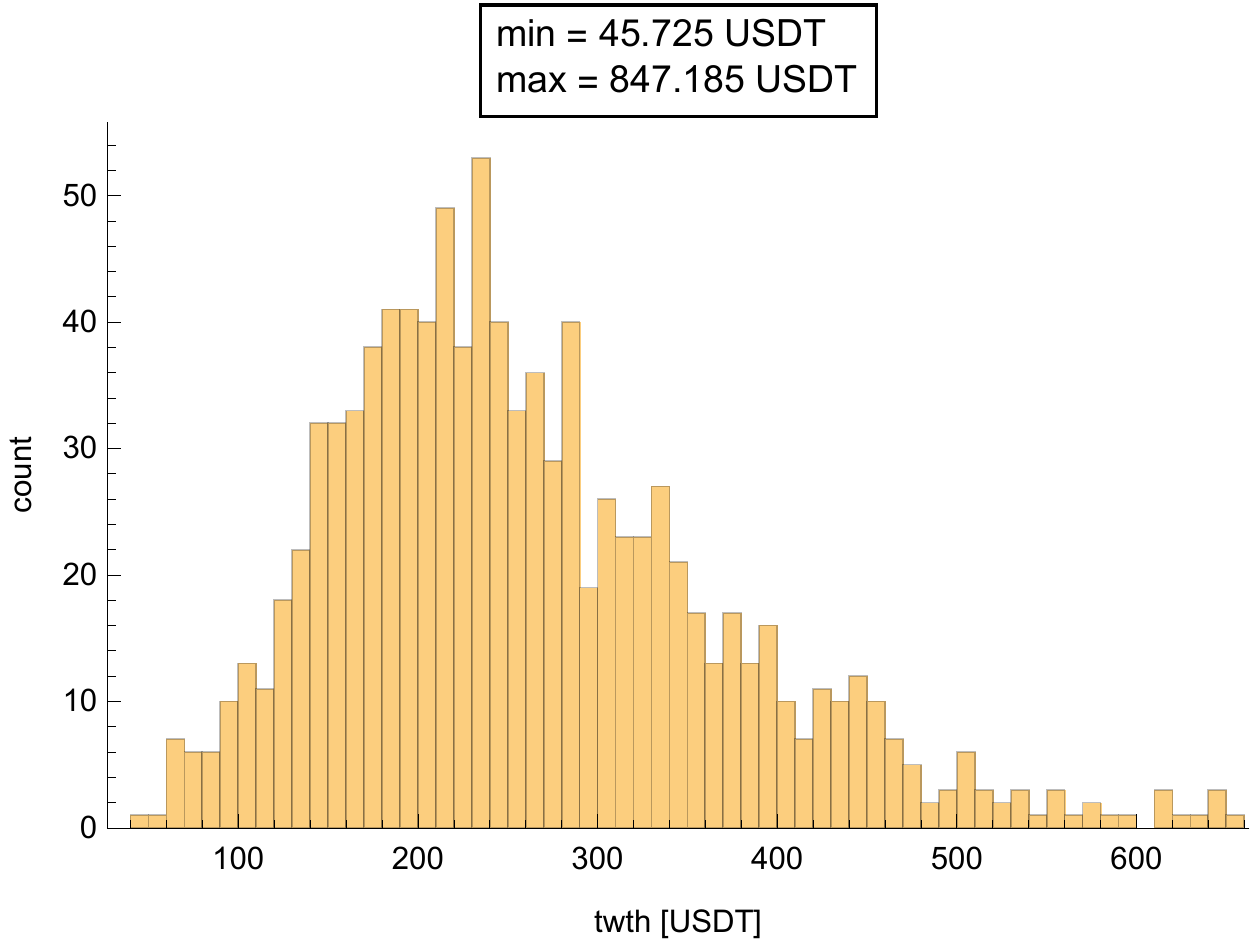}
\caption{$twth$ when taking non-random actions.}
\end{subfigure}
\hspace{0.8 mm}
\begin{subfigure}[t]{0.32\textwidth}
\includegraphics[width=\textwidth]{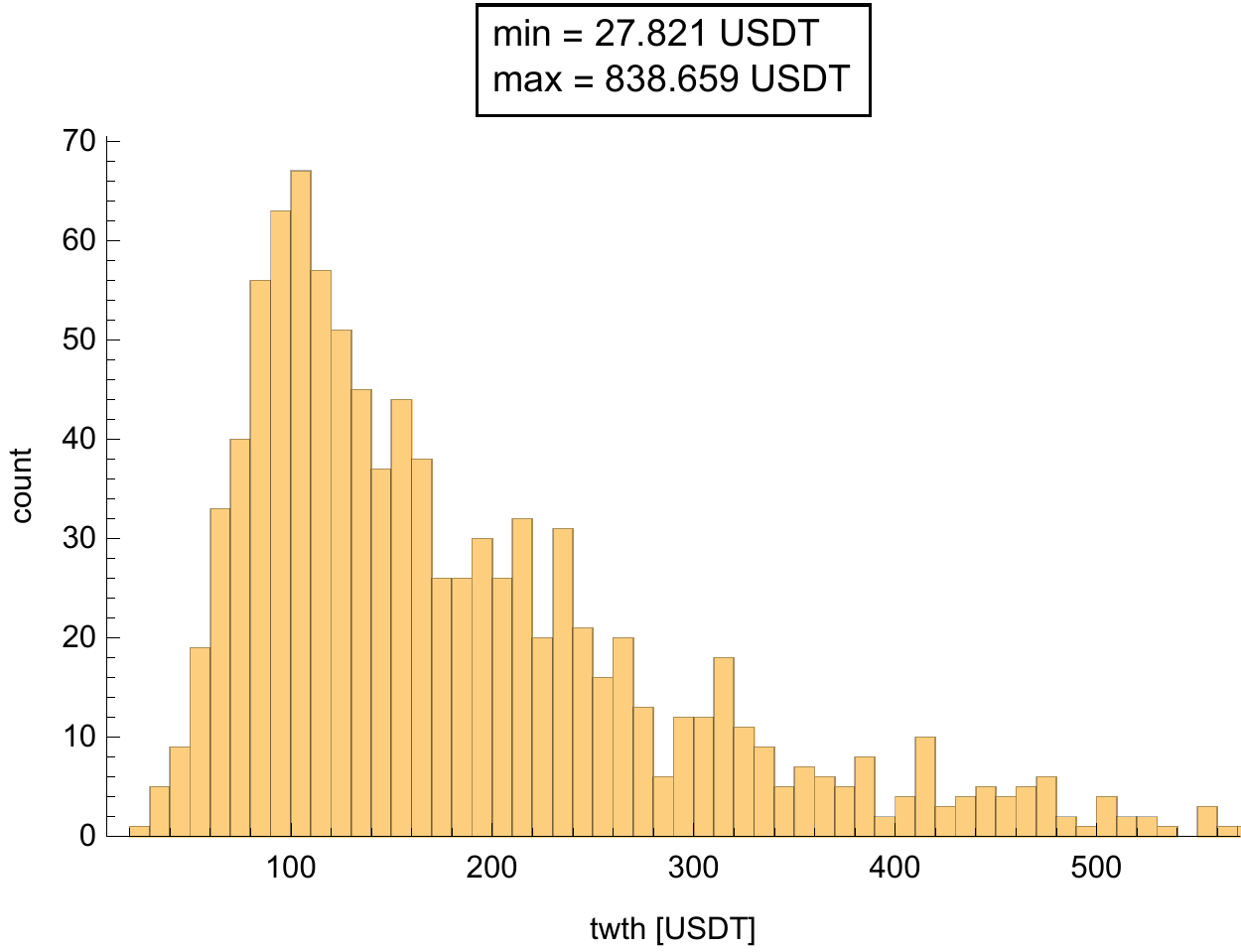}
\caption{$twth$ when taking random actions.}
\end{subfigure}
\hspace{0.8 mm}
\begin{subfigure}[t]{0.32\textwidth}
\includegraphics[width=\textwidth]{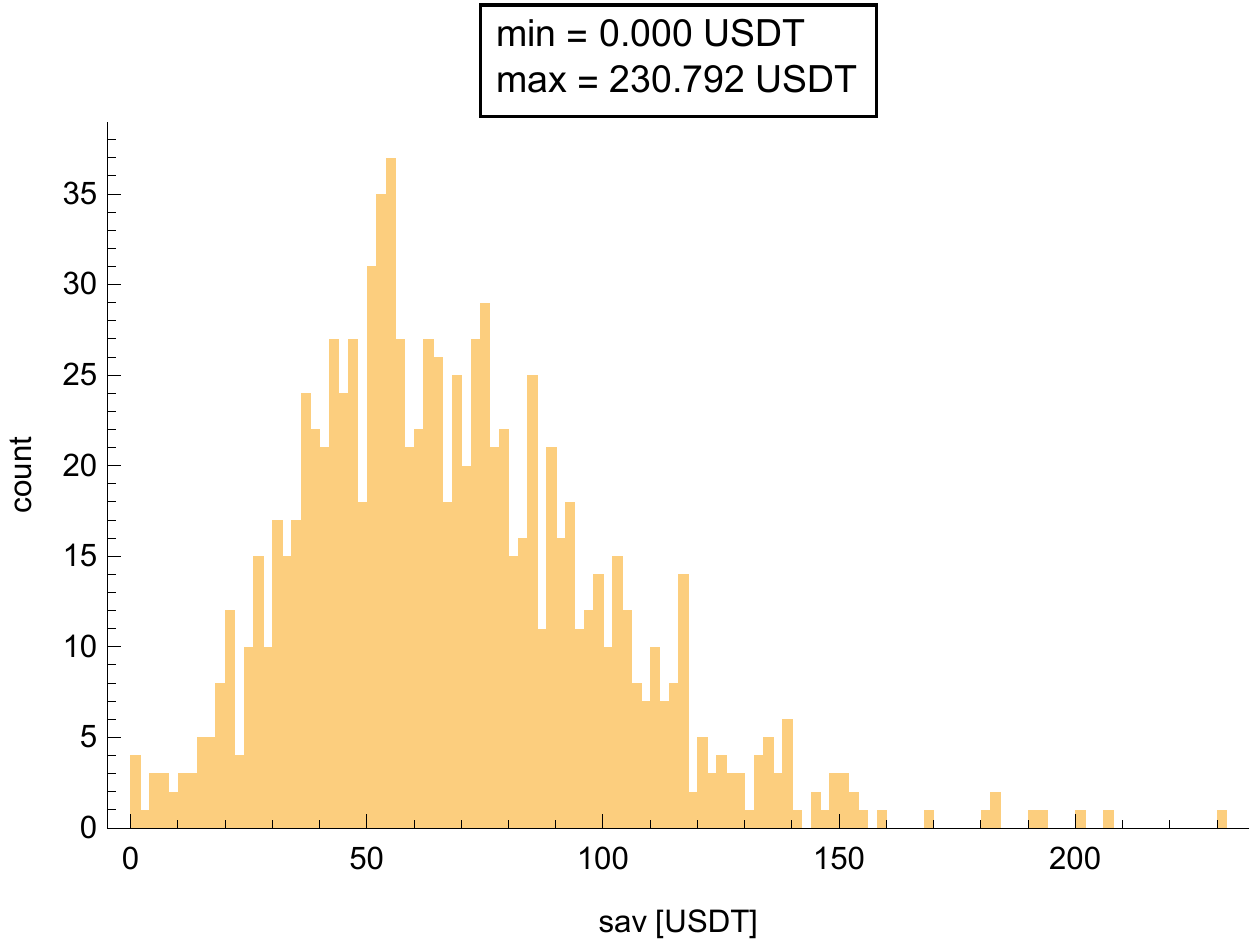}
\caption{$sav$ (when taking non-random actions).}
\end{subfigure}
\caption{Histograms for the full dataset.}\label{hist_full}
\end{figure}

\begin{figure}[H]
\centering
\begin{subfigure}[t]{0.32\textwidth}
\includegraphics[width=\textwidth]{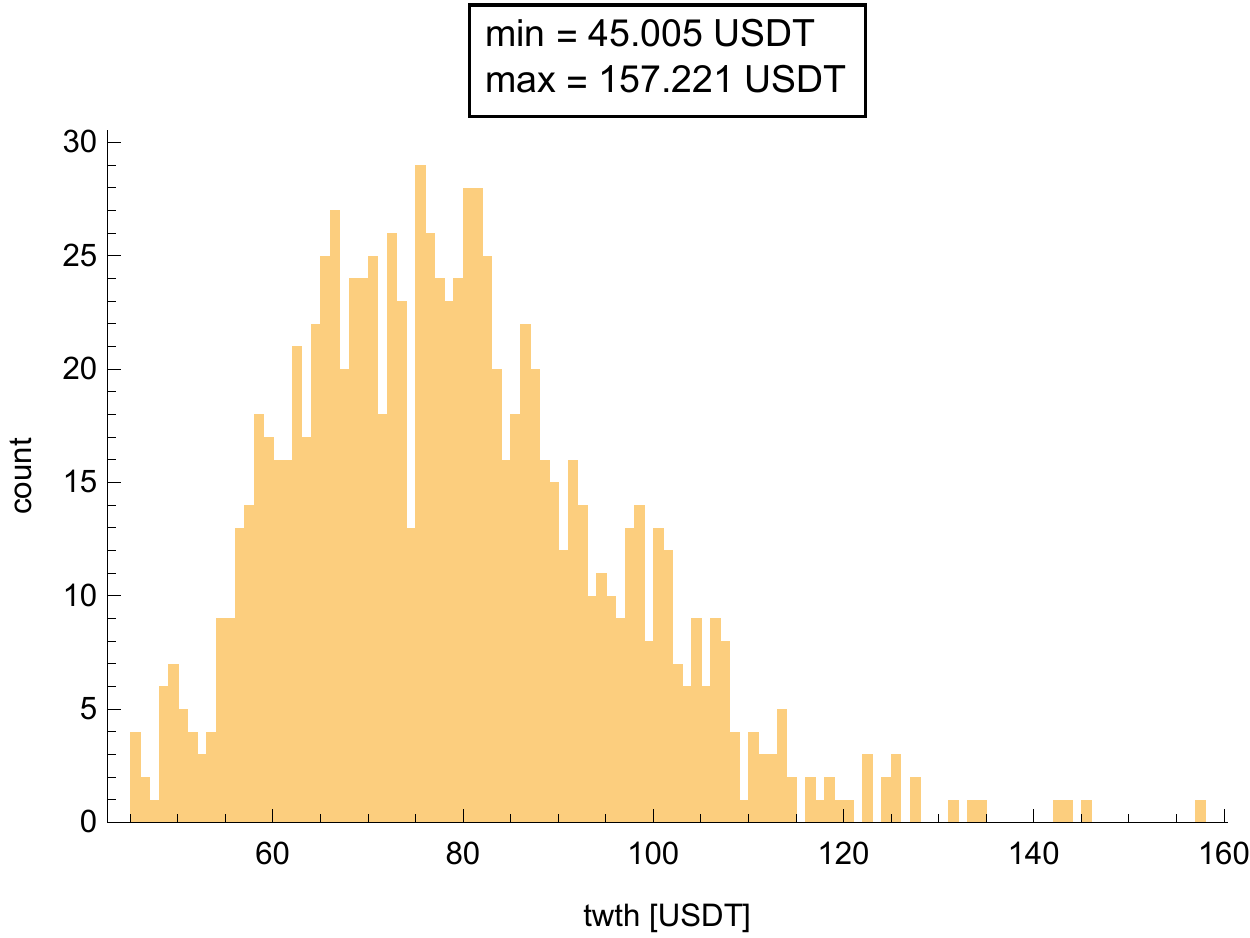}
\caption{$twth$ when taking non-random actions.}
\end{subfigure}
\hspace{0.8 mm}
\begin{subfigure}[t]{0.32\textwidth}
\includegraphics[width=\textwidth]{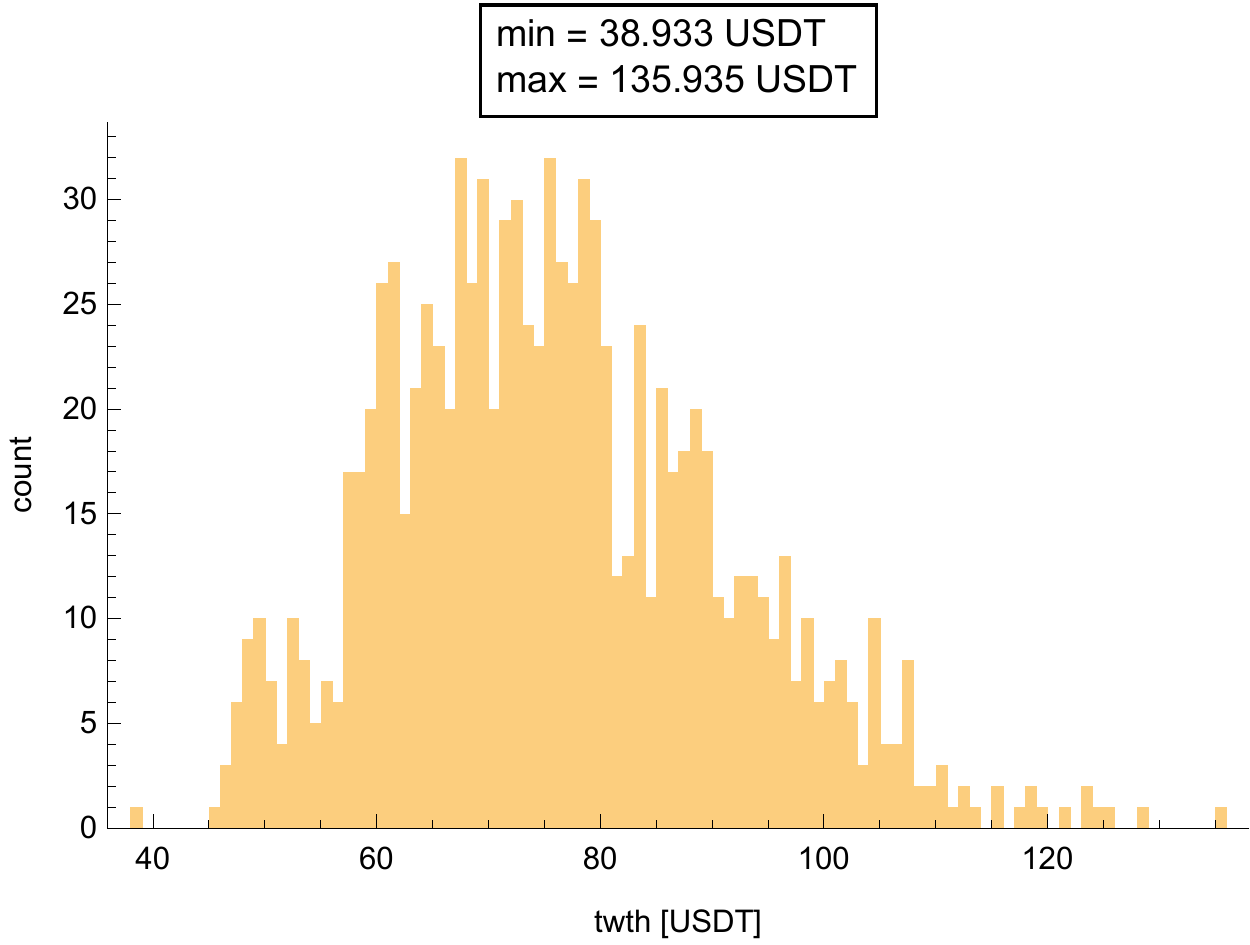}
\caption{$twth$ when taking random actions.}
\end{subfigure}
\hspace{0.8 mm}
\begin{subfigure}[t]{0.32\textwidth}
\includegraphics[width=\textwidth]{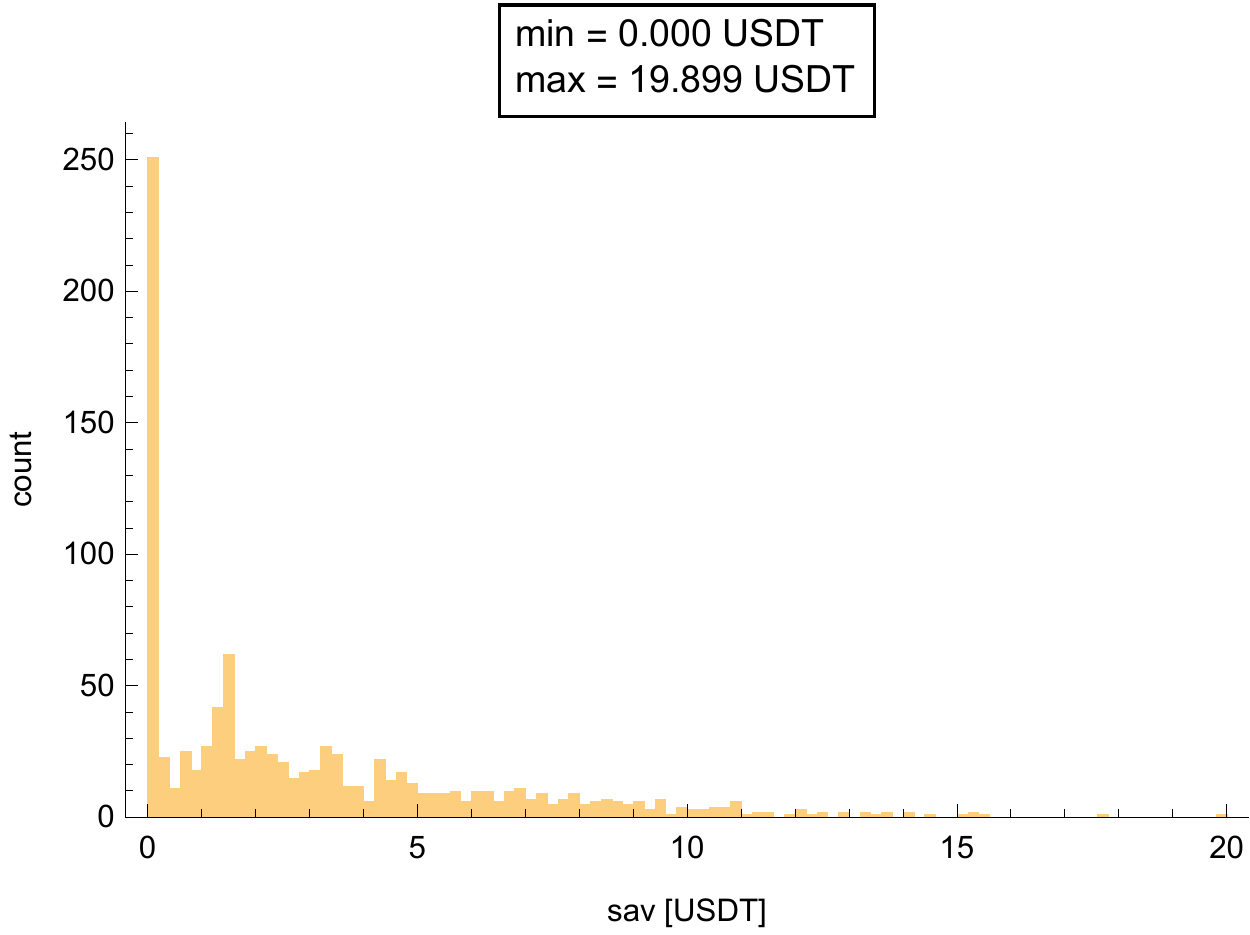}
\caption{$sav$ (when taking non-random actions).}
\end{subfigure}
\caption{Histograms for the ``bearish'' dataset.}\label{hist_bear}
\end{figure}

\begin{figure}[H]
\centering
\begin{subfigure}[t]{0.32\textwidth}
\includegraphics[width=\textwidth]{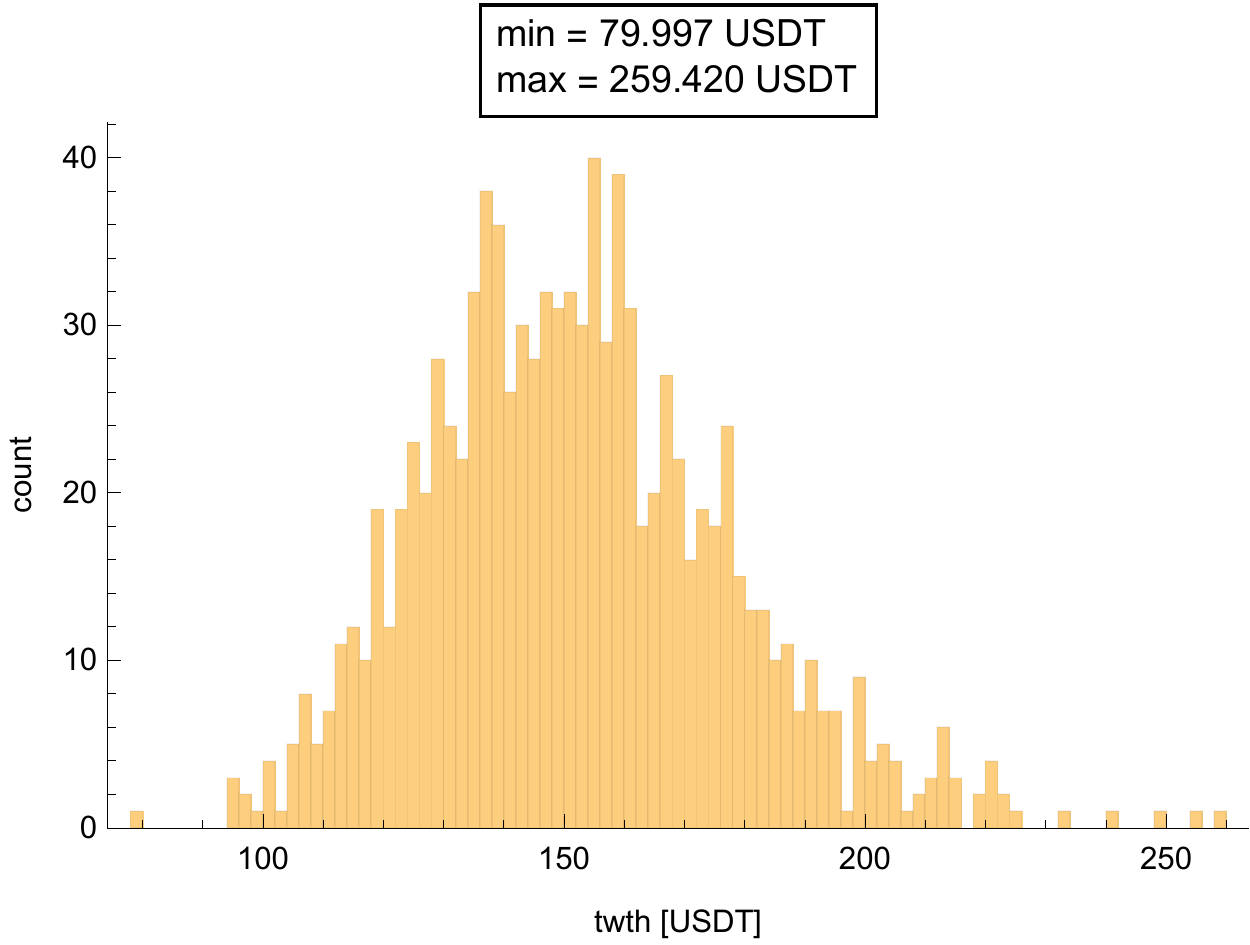}
\caption{$twth$ when taking non-random actions.}
\end{subfigure}
\hspace{0.8 mm}
\begin{subfigure}[t]{0.32\textwidth}
\includegraphics[width=\textwidth]{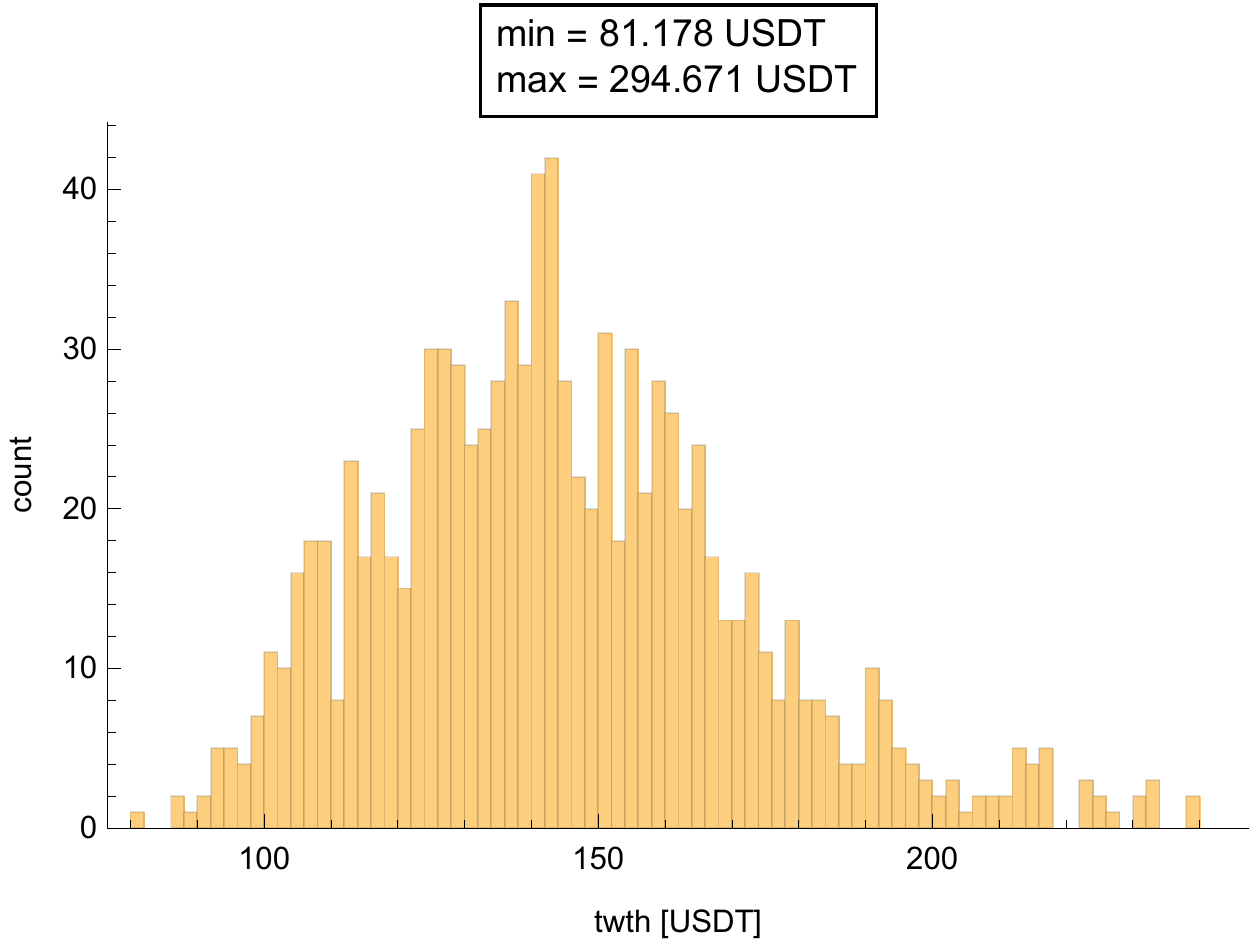}
\caption{$twth$ when taking random actions.}
\end{subfigure}
\hspace{0.8 mm}
\begin{subfigure}[t]{0.32\textwidth}
\includegraphics[width=\textwidth]{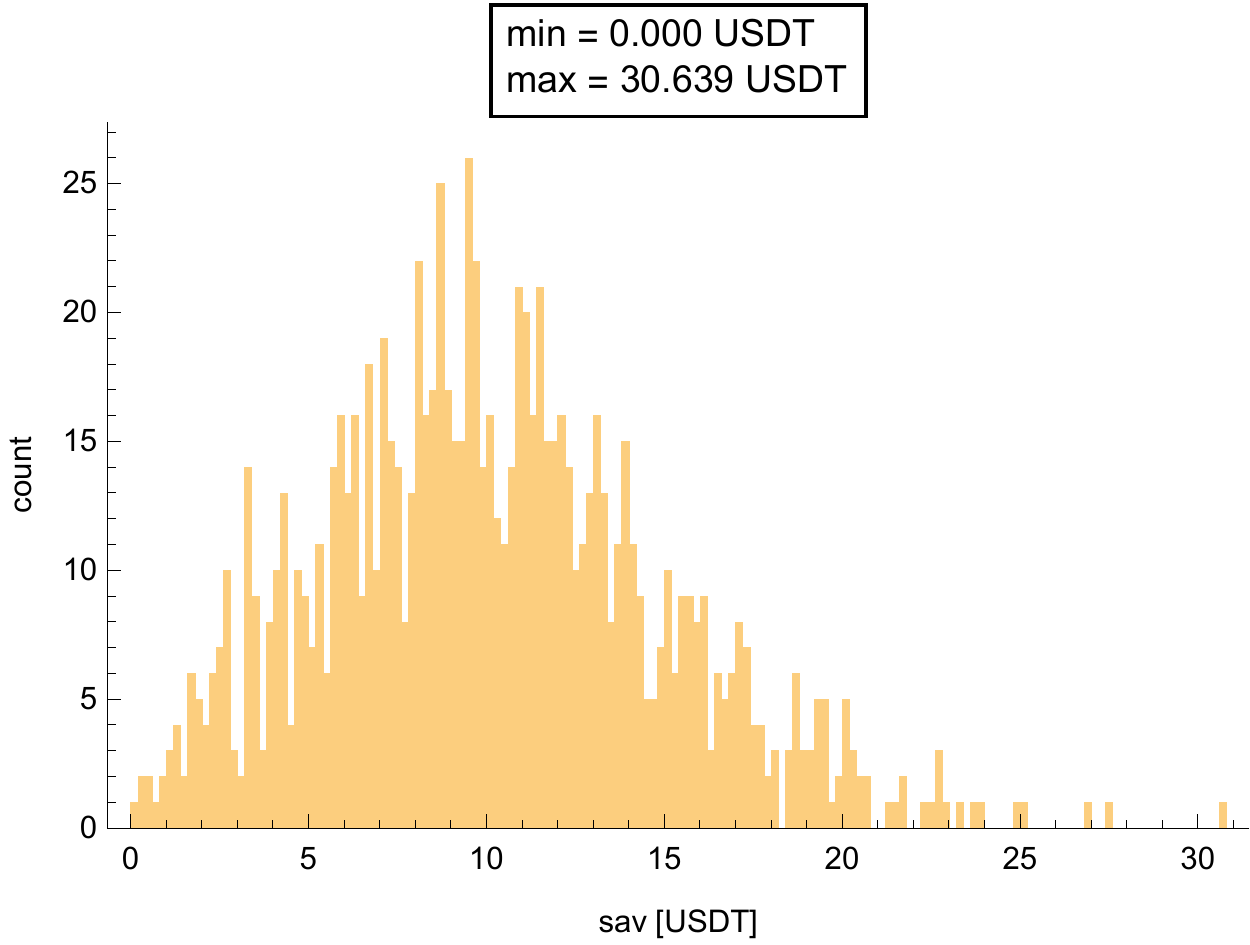}
\caption{$sav$ (when taking non-random actions).}
\end{subfigure}
\caption{Histograms for the ``bullish'' dataset.}\label{hist_bull}
\end{figure}

\begin{figure}[H]
\centering
\begin{subfigure}[t]{0.32\textwidth}
\includegraphics[width=\textwidth]{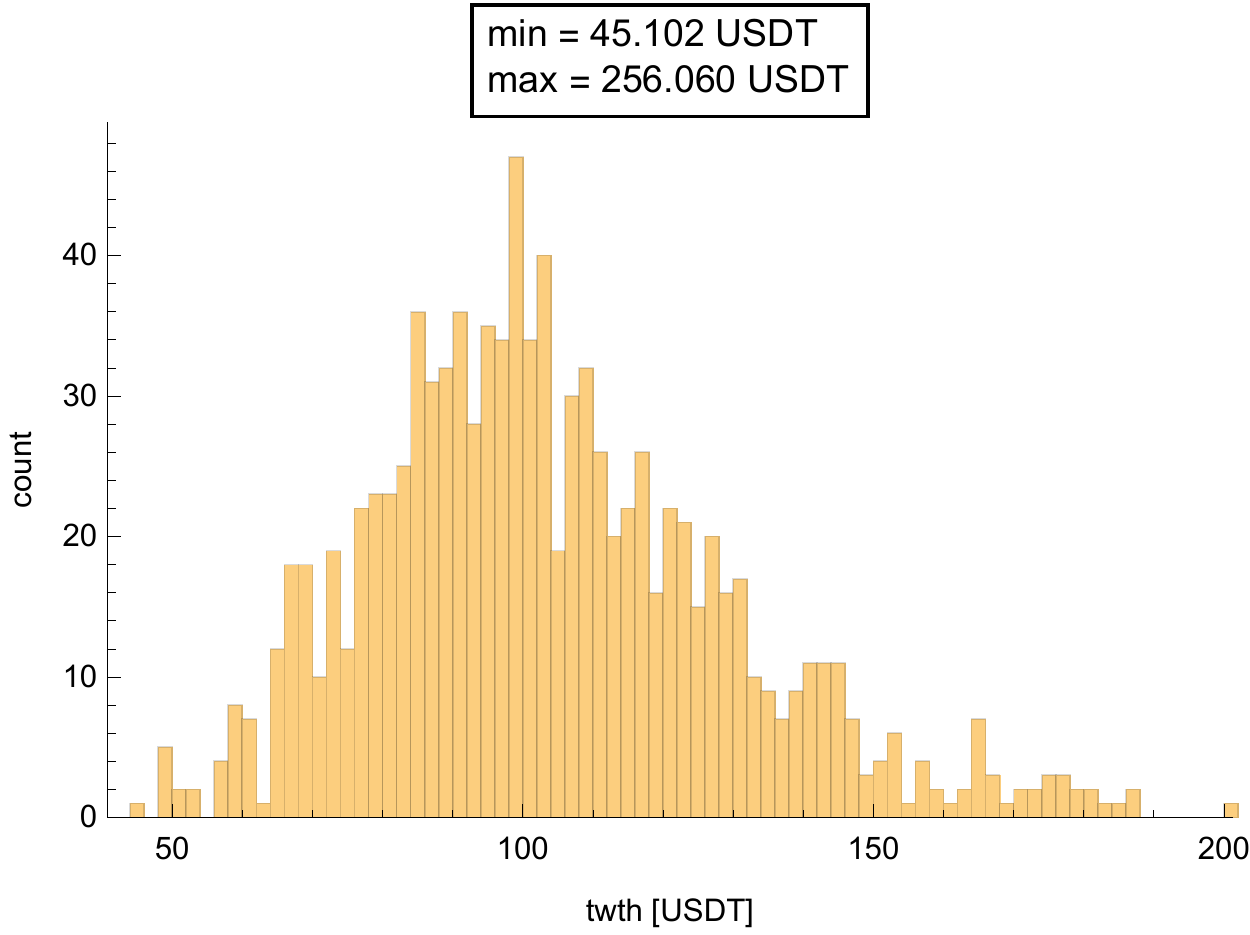}
\caption{$twth$ when taking non-random actions.}
\end{subfigure}
\hspace{0.8 mm}
\begin{subfigure}[t]{0.32\textwidth}
\includegraphics[width=\textwidth]{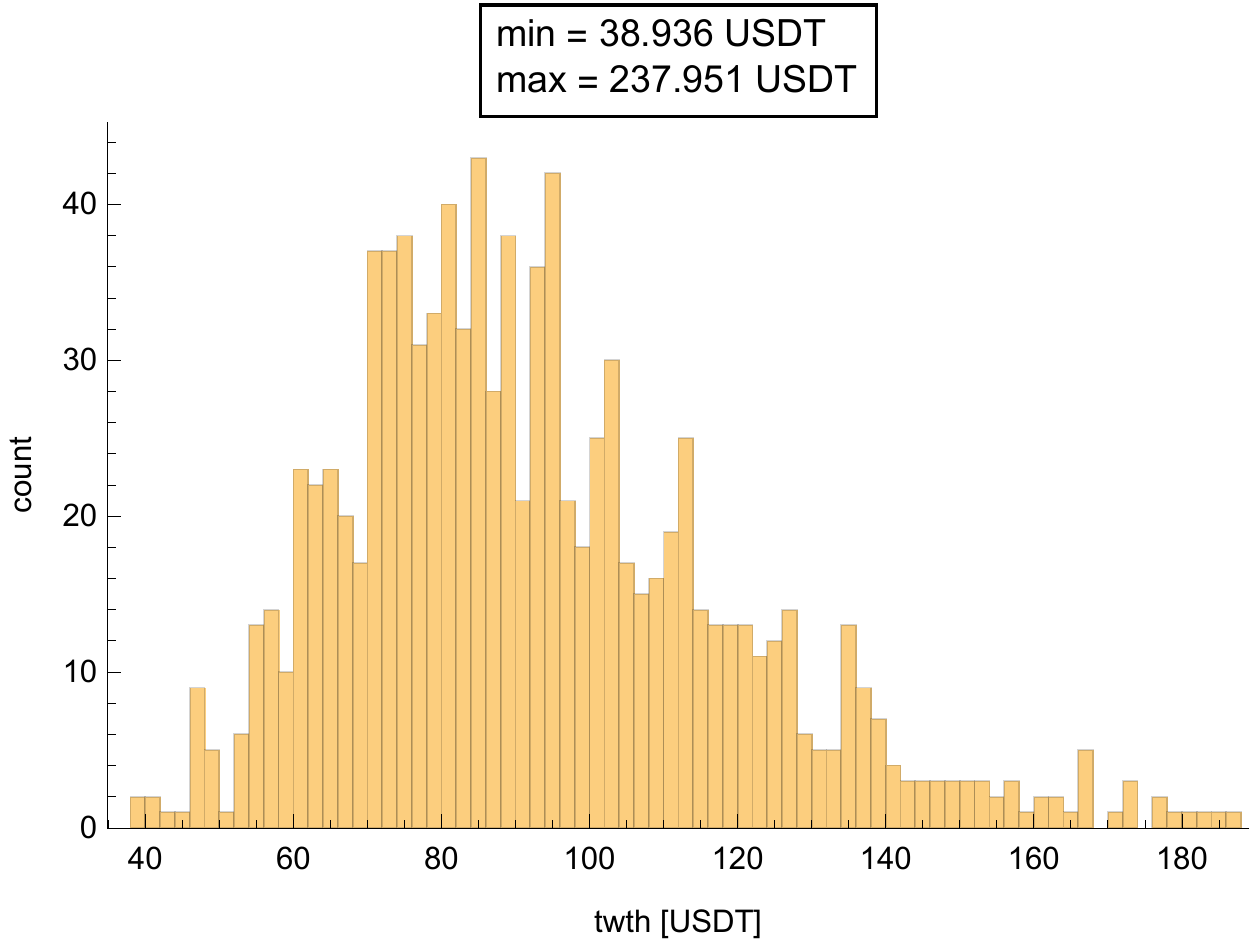}
\caption{$twth$ when taking random actions.}
\end{subfigure}
\hspace{0.8 mm}
\begin{subfigure}[t]{0.32\textwidth}
\includegraphics[width=\textwidth]{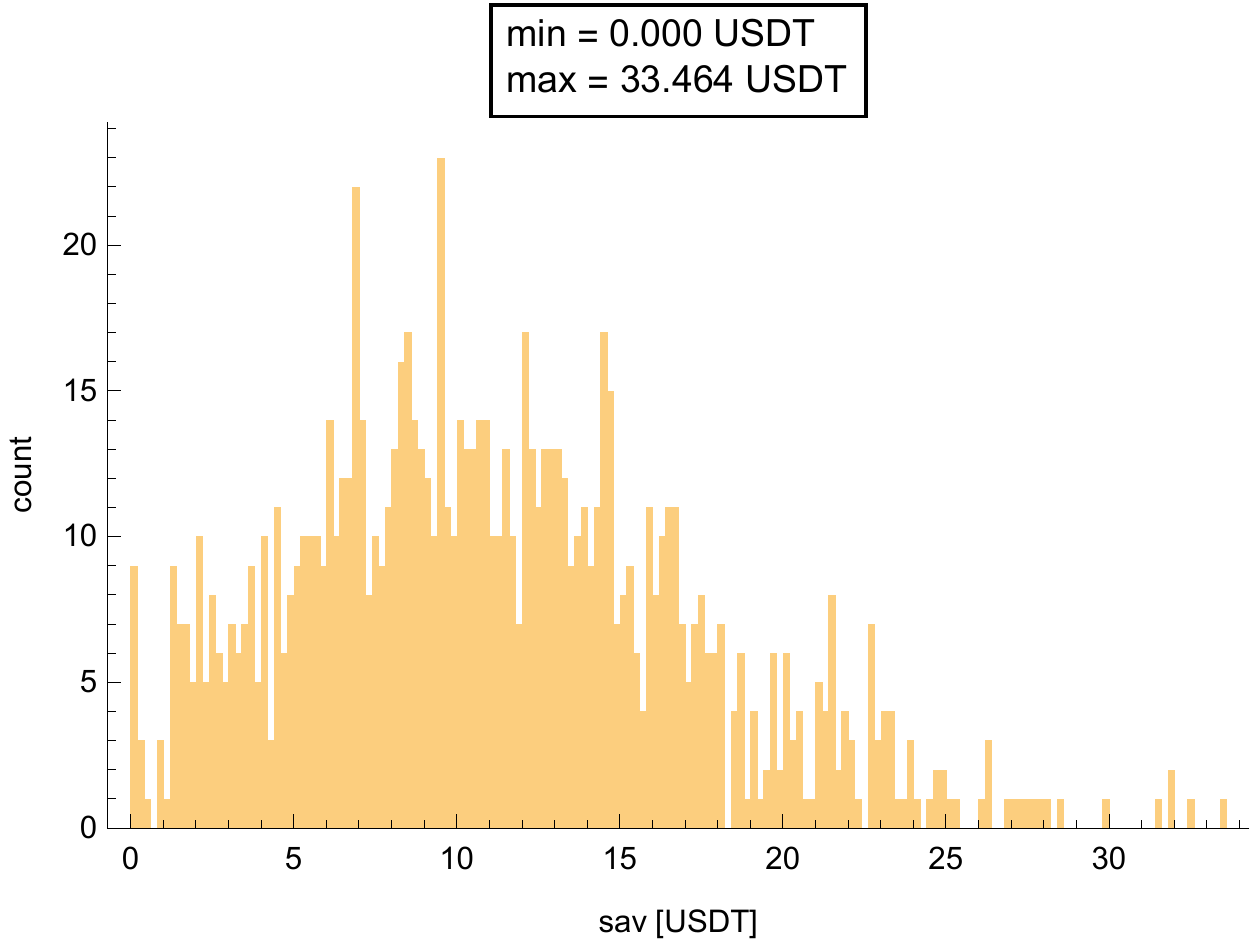}
\caption{$sav$ (when taking non-random actions).}
\end{subfigure}
\caption{Histograms for the ``mixed'' dataset.}\label{hist_mixed}
\end{figure}

\indent For each dataset we perform $1000$ runs of the algorithm.  Each run starts with $mon=100$, $cns=0$, $sav=0$, $res=0$ and $mlim=mon$ and we record the performance in terms of the sum of the wealth (\ref{wth}) and the $sav$ and $res$ pools, i.e.
\begin{align}
twth=wth+sav+res,
\end{align}

\noindent as well as the value of $sav$ alone, at the end of the run. In order to evaluate the effectiveness of the algorithm we also perform $1000$ runs with randomly selected actions on each time step. In this case $twth=wth$, as the savings mechanism depends on multiple reinforcement learning ingredients.\\
\indent After this we arrange the data in histograms (figures \ref{hist_full}, \ref{hist_bear}, \ref{hist_bull} and \ref{hist_mixed}), which also include the sample minimum and maximum, and calculate the sample means, medians, standard deviations, as well as the empirical probabilities for finishing a run with $twth\le 100$. The last is included as a measure of the risk of losing money after a run. All of these are arranged in tables \ref{tab_full}, \ref{tab_bear}, \ref{tab_bull} and \ref{tab_mixed}. As can be seen, our algorithm performs better than random in all datasets.\\
\indent In particular, for the full dataset we observe an increase in the mean value of $twth$ of about $39\%$ when taking non-random actions versus random ones. The median also increases -- by $58\%$. Additionally, the probability for finishing a run with $twth\le 100$ (i.e. for losing money) is $86\%$ smaller when taking non-random actions.
\begin{table}[H]
\centering
\begin{tabular}{ |c||c|c|c|c| }
\hline
& Mean [USDT] & Median [USDT] & St. Dev. [USDT] & $P(twth\le 100)$  \\
\hline\hline
Non-random $twth$ & 263.928 & 241.903 & 113.334 & 0.031 \\ 
\hline
Random $twth$ & 189.703 & 153.028 & 121.777 & 0.226 \\
\hline
$sav$ & 68.192 & 64.020 & 32.131 & $-$ \\
\hline
\end{tabular}
\caption{Sample means, medians and standard deviations of $twth$ when taking random or non-random actions and of $sav$ for the full dataset. The probability for finishing a run with $twth\le 100$ is also included.}
\label{tab_full}
\end{table}

\indent Similar calculations can be made for the rest of the datasets. In all of the cases the algorithm has higher mean and median values of $twth$ when compared to random, as well as lower values of $P(twth\le 100)$, i.e. it makes more money on average and has a lower chance to lose money. Probably the most interesting case is the ``bearish'' market one, as there it is the hardest to make a profit. This is reflected in our results as the differences with random are the smallest. More specifically, the mean and median of $twth$ are about $4\%$ larger and the probability of $twth\le 100$ -- about $5\%$ smaller.
\begin{table}[H]
\centering
\begin{tabular}{ |c||c|c|c|c| }
\hline
& Mean [USDT] & Median [USDT] & St. Dev. [USDT] & $P(twth\le 100)$  \\
\hline\hline
Non-random $twth$ & $78.999$ & $77.538$ & $16.689$ & $0.884$ \\ 
\hline
Random $twth$ & $76.139$ & $74.905$ & $15.169$ & $0.926$ \\
\hline
$sav$ & $3.060$ & $1.982$ & $3.324$ & $-$ \\
\hline
\end{tabular}
\caption{Descriptive statistics of $twth$ when taking random or non-random actions and of $sav$ for the ``bearish'' dataset.}
\label{tab_bear}
\end{table}

\indent For completeness we also include the relative performance in the other two datasets. In terms of the mean of $twth$ our algorithm performs about $5\%$ better in the ``bullish'' dataset and $12\%$ better in the ``mixed'' dataset. In terms of the median the increases are $6\%$ and $14\%$ for the ``bullish'' and ``mixed'' cases, respectively. Finally, in terms of $P(twth\le 100)$ the decreases are $74\%$ and $26\%$, respectively, for the two cases.
\begin{table}[H]
\centering
\begin{tabular}{ |c||c|c|c|c| }
\hline
& Mean [USDT] & Median [USDT] & St. Dev. [USDT] & $P(twth\le 100)$  \\
\hline\hline
Non-random $twth$ & $152.267$ & $150.601$ & $25.372$ & $0.007$ \\ 
\hline
Random $twth$ & $145.245$ & $142.256$ & $28.561$ & $0.027$ \\
\hline
$sav$ & $10.169$ & $9.758$ & $4.755$ & $-$\\
\hline
\end{tabular}
\caption{Descriptive statistics for the ``bullish'' dataset.}
\label{tab_bull}
\end{table}

\begin{table}[H]
\centering
\begin{tabular}{ |c||c|c|c|c| }
\hline
& Mean [USDT] & Median [USDT] & St. Dev. [USDT] & $P(twth\le 100)$  \\
\hline\hline
Non-random $twth$ & $104.338$ & $100.545$ & $26.771$ & $0.491$ \\ 
\hline
Random $twth$ & $93.202$ & $88.569$ & $26.994$ & $0.664$ \\
\hline
$sav$ & $11.182$ & $10.610$ & $6.039$ & $-$\\
\hline
\end{tabular}
\caption{Descriptive statistics for the ``mixed'' dataset.}
\label{tab_mixed}
\end{table}

\section{Conclusion}\label{conc}

In this paper we attempted to tackle the challenging problem of market prediction using machine learning. More specifically, we introduced a deep reinforcement learning agent that is meant to adapt to market conditions and trade fully online. The main components of our algorithm are a more or less standard Double $Q$-learning framework coupled with a Fast Learning Network, used to approximate the $Q$-functions. On top of that we added a mechanism, which takes money out of the trading pool, both as a means to take profit and to boost performance by reusing some of it at a more favourable moment.\\
\indent After this we tested the algorithm on historical market data, which was chosen so that it captures different market conditions. We observed that our agent performs better than random on all datasets -- both in terms of profit and probability of loss at the end of a run through the data. Furthermore, it did so even in a ``bearish'' market, when the overall market trend is downward and, most importantly, without any prior learning on big offline dataset. We can view the latter as the main strength of our algorithm.

\newpage
\begin{appendices}

\section{Mathematica code}\label{code}
\tiny{\noindent\(\pmb{\text{CloseKernels}[];}\\
\pmb{\text{LaunchKernels}[];}\\
\pmb{\text{ClearAll}[\text{{``}Global$\grave{ }$*{''}}]}\\
\pmb{\text{SetDirectory}[\text{NotebookDirectory}[]]}\\
\pmb{}\\
\pmb{\text{(* Read the CSV file with the prices and volumes. *)}}\\
\pmb{\text{data}=\text{Import}[\text{{``}ADA-USDT.csv{''}}];}\\
\pmb{\text{uprices}=\text{Table}[\text{data}[[k,2]],\{k,2,\text{Length}[\text{data}]\}];\text{simlength}=\text{Length}[\text{uprices}]}\\
\pmb{\text{uvolumes}=\text{Table}[\text{data}[[k,6]]/10000000,\{k,2,\text{Length}[\text{data}]\}];}\\
\pmb{}\\
\pmb{\text{(* Pass the prices through a filer and make a list of the new prices and preceding volumes. *)}}\\
\pmb{\text{prices}=\text{Table}[0,\{k,1,\text{simlength}\}];}\\
\pmb{\text{volumes}=\text{Table}[0,\{k,1,\text{simlength}\}];}\\
\pmb{\text{pinit}=\text{uprices}[[2]];}\\
\pmb{\text{prices}[[1]]=\text{uprices}[[2]];}\\
\pmb{\text{volumes}[[1]]=\text{uvolumes}[[1]];}\\
\pmb{l=2;}\\
\pmb{i=3;}\\
\pmb{\text{While}[i\leq \text{simlength},}\\
\pmb{\text{If}[\text{Abs}[\text{uprices}[[i]]-\text{pinit}]/\text{pinit}>0.01,\text{pinit}=\text{uprices}[[i]];\text{prices}[[l]]=\text{uprices}[[i]];\text{volumes}[[l]]=\text{uvolumes}[[i-1]];l=l+1];}\\
\pmb{i=i+1}\\
\pmb{];}\\
\pmb{\text{prices}=\text{DeleteCases}[\text{prices},0];}\\
\pmb{\text{simlength}=\text{Length}[\text{prices}]}\\
\pmb{\text{volumes}=\text{DeleteCases}[\text{volumes},0];}\\
\pmb{}\\
\pmb{\text{(* Plot the prices and volumes. *)}}\\
\pmb{\text{ListPlot}[\text{prices},\text{Joined}\to \text{True},\text{PlotRange}\to \text{Full}]}\\
\pmb{\text{ListPlot}[\text{volumes},\text{Joined}\to \text{True},\text{PlotRange}\to \text{Full}]}\\
\pmb{}\\
\pmb{\text{(* Parameters. *)}}\\
\pmb{\text{vsize}=5;\text{(* Number of data points for each state. *)}}\\
\pmb{\text{fsize}=1+\text{vsize}+1+1+1+1+1+1+1+1+1+1+4+3+2+1+1\text{(* Size of the feature vector. *)}}\\
\pmb{\text{hlsize}=50;\text{(* Size of the hidden layer. *)}}\\
\pmb{\text{gamma}=0.05;\text{(* Discount factor. *)}}\\
\pmb{\text{probeps}=0.0001;\text{(* Probability to reset the value of epsilon. *)}}\\
\pmb{\text{mlimn}=75.;\text{(* The lowest possible value of mlim. *)}}\\
\pmb{\text{runs}=1000;\text{(* Number of test runs. *)}}\\
\pmb{}\\
\pmb{\text{feat}=\text{Table}[0.,\{k,1,\text{fsize}\}];\text{(* Feature vector. *)}}\\
\pmb{\text{pr}=\text{Table}[0.,\{k,1,\text{vsize}\}];\text{(* List of prices in the state. *)}}\\
\pmb{}\\
\pmb{\text{(* Hidden layer outputs. *)}}\\
\pmb{\text{hlayer1}=\text{Table}[0.,\{k,1,\text{hlsize}\}];}\\
\pmb{\text{hlayer2}=\text{Table}[0.,\{k,1,\text{hlsize}\}];}\\
\pmb{}\\
\pmb{\text{(* Gradients of the Q-functions. *)}}\\
\pmb{\text{gradq1}=\text{Table}[0.,\{k,1,\text{hlsize}+\text{fsize}\}];}\\
\pmb{\text{gradq1new}=\text{Table}[0.,\{k,1,\text{hlsize}+\text{fsize}\}];}\\
\pmb{\text{gradq2}=\text{Table}[0.,\{k,1,\text{hlsize}+\text{fsize}\}];}\\
\pmb{\text{gradq2new}=\text{Table}[0.,\{k,1,\text{hlsize}+\text{fsize}\}];}\\
\pmb{}\\
\pmb{\text{(* Q-functions. *)}}\\
\pmb{\text{qarr1}=\text{Table}[0.,\{k,1,19\}];}\\
\pmb{\text{qarr2}=\text{Table}[0.,\{k,1,19\}];}\\
\pmb{\text{qarr}=\text{Table}[0.,\{k,1,19\}];}\\
\pmb{\text{qarr1new}=\text{Table}[0.,\{k,1,19\}];}\\
\pmb{\text{qarr2new}=\text{Table}[0.,\{k,1,19\}];}\\
\pmb{}\\
\pmb{\text{(* Lists to write sav and twth at the end of each run. *)}}\\
\pmb{\text{savarray}=\text{Table}[0.,\{k,1,\text{runs}\}];}\\
\pmb{\text{twtharray}=\text{Table}[0.,\{k,1,\text{runs}\}];}\\
\pmb{}\\
\pmb{\text{(* Relative movements of the price, etc. *)}}\\
\pmb{\text{nmd1}=\text{Table}[0.,\{k,1,\text{vsize}-1\}];}\\
\pmb{\text{nmd2}=\text{Table}[0.,\{k,1,\text{vsize}-2\}];}\\
\pmb{\text{nmd3}=\text{Table}[0.,\{k,1,\text{vsize}-3\}];}\\
\pmb{\text{nmd4}=\text{Table}[0.,\{k,1,\text{vsize}-4\}];}\\
\pmb{}\\
\pmb{\text{(* The activation function. *)}}\\
\pmb{\text{Plot}[1./(1.+\text{Exp}[-x]),\{x,-6.,6.\}]}\\
\pmb{}\\
\pmb{m=1;\text{While}[m\leq \text{runs},}\\
\pmb{}\\
\pmb{\text{(* Initial weights are randomly generated. Then the weights between the input and hidden layers are rescaled. *)}}\\
\pmb{\text{whi1}=\text{Table}[\text{RandomReal}[\{-1.,1.\}],\{k,1,\text{hlsize}\},\{l,1,\text{fsize}\}];}\\
\pmb{j=1;\text{While}[j\leq \text{hlsize},\text{whi1}[[j]]=\text{whi1}[[j]]/\text{Norm}[\text{whi1}[[j]]];j=j+1];}\\
\pmb{\text{whi2}=\text{Table}[\text{RandomReal}[\{-1.,1.\}],\{k,1,\text{hlsize}\},\{l,1,\text{fsize}\}];}\\
\pmb{j=1;\text{While}[j\leq \text{hlsize},\text{whi2}[[j]]=\text{whi2}[[j]]/\text{Norm}[\text{whi2}[[j]]];j=j+1];}\\
\pmb{\text{wout1}=\text{Table}[\text{RandomReal}[\{-1.,1.\}],\{l,1,19\},\{k,1,\text{hlsize}+\text{fsize}\}];}\\
\pmb{\text{wout2}=\text{Table}[\text{RandomReal}[\{-1.,1.\}],\{l,1,19\},\{k,1,\text{hlsize}+\text{fsize}\}];}\\
\pmb{}\\
\pmb{\text{(* Initial values of the counters for epsilon and alpha. *)}}\\
\pmb{\text{ialpha}=-1;}\\
\pmb{\text{ieps}=0;}\\
\pmb{}\\
\pmb{\text{(* Tables to be used for calculating RSI. *)}}\\
\pmb{\text{rsip}=\text{Table}[0.,\{l,1,15\}];}\\
\pmb{\text{rsiu}=\text{Table}[0.,\{l,1,14\}];}\\
\pmb{\text{rsid}=\text{Table}[0.,\{l,1,14\}];}\\
\pmb{}\\
\pmb{\text{(* Variables for the average volumes and the max norms of the output weights. *)}}\\
\pmb{\text{mv}=\text{Table}[0.,\{j,1,20\}];}\\
\pmb{\text{nv}=0;}\\
\pmb{\text{av}=0.;}\\
\pmb{\text{max1}=1.;}\\
\pmb{\text{max2}=1.;}\\
\pmb{}\\
\pmb{\text{(* Initial mon, cns, sav, res, mlim. *)}}\\
\pmb{\text{mon}=100.;\text{cns}=0.; \text{sav}=0.;\text{res}=0.;\text{mlim}=\text{mon};\text{mdf}=0;}\\
\pmb{}\\
\pmb{i=1;}\\
\pmb{}\\
\pmb{\text{Label}[\text{episode}];\text{(* Label used to start a new episode. *)}}\\
\pmb{}\\
\pmb{\text{(* Initial price in the episode. *)}}\\
\pmb{\text{ipr}=\text{prices}[[i*5-\text{vsize}+1]];}\\
\pmb{}\\
\pmb{\text{(* Calculate the average volumes. *)}}\\
\pmb{\text{cav}=(\text{volumes}[[i*5-4]]+\text{volumes}[[i*5-3]]+\text{volumes}[[i*5-2]]+\text{volumes}[[i*5-1]]+\text{volumes}[[i*5]])/5.;}\\
\pmb{j=1;\text{While}[j\leq 19,\text{mv}[[j]]=\text{mv}[[j+1]];j=j+1];}\\
\pmb{\text{mv}[[j]]=\text{cav};}\\
\pmb{\text{av}=\text{Mean}[\text{mv}];}\\
\pmb{}\\
\pmb{\text{(* Observe the prices in the current state. *)}}\\
\pmb{j=1;\text{While}[j\leq \text{vsize},\text{pr}[[j]]=\text{prices}[[i*5-\text{vsize}+j]];j=j+1];}\\
\pmb{}\\
\pmb{\text{(* Relative price movements. *)}}\\
\pmb{j=1;\text{While}[j\leq \text{vsize}-1,}\\
\pmb{\text{nmd1}[[j]]=(\text{pr}[[j+1]]-\text{pr}[[j]])/\text{pr}[[j]];}\\
\pmb{j=j+1];}\\
\pmb{j=1;\text{While}[j\leq \text{vsize}-2,}\\
\pmb{\text{nmd2}[[j]]=(\text{nmd1}[[j+1]]-\text{nmd1}[[j]])/\text{Abs}[\text{nmd1}[[j]]];}\\
\pmb{j=j+1];}\\
\pmb{j=1;\text{While}[j\leq \text{vsize}-3,}\\
\pmb{\text{nmd3}[[j]]=(\text{nmd2}[[j+1]]-\text{nmd2}[[j]])/\text{Abs}[\text{nmd2}[[j]]];}\\
\pmb{j=j+1];}\\
\pmb{j=1;\text{While}[j\leq \text{vsize}-4,}\\
\pmb{\text{nmd4}[[j]]=(\text{nmd3}[[j+1]]-\text{nmd3}[[j]])/\text{Abs}[\text{nmd3}[[j]]];}\\
\pmb{j=j+1];}\\
\pmb{}\\
\pmb{\text{(* Calculate RSI for 15 points. *)}}\\
\pmb{j=1;\text{While}[j\leq 10,\text{rsip}[[j]]=\text{rsip}[[j+5]];j=j+1];}\\
\pmb{j=1;\text{While}[j\leq \text{vsize},\text{rsip}[[10+j]]=\text{pr}[[j]];j=j+1];}\\
\pmb{j=1;\text{While}[j\leq 14,}\\
\pmb{\text{If}[\text{rsip}[[j+1]]>\text{rsip}[[j]],\text{rsiu}[[j]]=\text{rsip}[[j+1]]-\text{rsip}[[j]];\text{rsid}[[j]]=0.];}\\
\pmb{\text{If}[\text{rsip}[[j+1]]<\text{rsip}[[j]],\text{rsid}[[j]]=\text{rsip}[[j]]-\text{rsip}[[j+1]];\text{rsiu}[[j]]=0.];}\\
\pmb{\text{If}[\text{rsip}[[j+1]]==\text{rsip}[[j]],\text{rsiu}[[j]]=0.;\text{rsid}[[j]]=0.];}\\
\pmb{j=j+1];}\\
\pmb{\text{If}[\text{Mean}[\text{rsid}]==0.,\text{rsi}=100.,\text{rsi}=100.-(100./(1.+(\text{Mean}[\text{rsiu}]/\text{Mean}[\text{rsid}])))];}\\
\pmb{}\\
\pmb{\text{(* Construct the feature vector (same for all actions). *)}}\\
\pmb{\text{feat}[[1]]=0.;}\\
\pmb{j=1;k=1;\text{While}[j\leq \text{vsize},\text{feat}[[k+1]]=\text{pr}[[j]];k=k+1;j=j+1];}\\
\pmb{\text{feat}[[k+1]]=\text{ipr};k=k+1;}\\
\pmb{\text{feat}[[k+1]]=(\text{pr}[[\text{vsize}]]-\text{ipr})/\text{ipr};k=k+1;}\\
\pmb{\text{feat}[[k+1]]=\text{mon};k=k+1;}\\
\pmb{\text{feat}[[k+1]]=\text{cns};k=k+1;}\\
\pmb{\text{feat}[[k+1]]=\text{cav};k=k+1;}\\
\pmb{\text{feat}[[k+1]]=\text{av};k=k+1;}\\
\pmb{\text{feat}[[k+1]]=(\text{cav}-\text{av})/\text{av};k=k+1;}\\
\pmb{\text{feat}[[k+1]]=(\text{volumes}[[i*5]]-\text{av})/\text{av};k=k+1;}\\
\pmb{\text{feat}[[k+1]]=(\text{volumes}[[i*5]]-\text{cav})/\text{cav};k=k+1;}\\
\pmb{\text{feat}[[k+1]]=\text{rsi};k=k+1;}\\
\pmb{j=1;\text{While}[j\leq \text{vsize}-1,\text{feat}[[k+1]]=\text{nmd1}[[j]];k=k+1;j=j+1];}\\
\pmb{j=1;\text{While}[j\leq \text{vsize}-2,\text{feat}[[k+1]]=\text{nmd2}[[j]];k=k+1;j=j+1];}\\
\pmb{j=1;\text{While}[j\leq \text{vsize}-3,\text{feat}[[k+1]]=\text{nmd3}[[j]];k=k+1;j=j+1];}\\
\pmb{j=1;\text{While}[j\leq \text{vsize}-4,\text{feat}[[k+1]]=\text{nmd4}[[j]];k=k+1;j=j+1];}\\
\pmb{\text{feat}[[k+1]]=\text{mlim};k=k+1;}\\
\pmb{}\\
\pmb{\text{(* Rescale the feature vector, so its norm is 6. *)}}\\
\pmb{\text{feat}=6.*(\text{feat}/\text{Norm}[\text{feat}]);}\\
\pmb{\text{feat}[[1]]=1.;}\\
\pmb{}\\
\pmb{\text{(* Output of the hidden layers. *)}}\\
\pmb{n=1;\text{While}[n\leq \text{hlsize},}\\
\pmb{q=0;j=1;\text{While}[j\leq \text{fsize},q=q+\text{whi1}[[n,j]]*\text{feat}[[j]];j=j+1];}\\
\pmb{\text{hlayer1}[[n]]=1./(1.+\text{Exp}[-q]);}\\
\pmb{q=0;j=1;\text{While}[j\leq \text{fsize},q=q+\text{whi2}[[n,j]]*\text{feat}[[j]];j=j+1];}\\
\pmb{\text{hlayer2}[[n]]=1./(1.+\text{Exp}[-q]);}\\
\pmb{n=n+1];}\\
\pmb{}\\
\pmb{\text{(* Gradients of the Q-functions, taken in the current state (same for all actions), which are just}}\\
\pmb{\text{concatenations of the feature vector with the hidden layer outputs. *)}}\\
\pmb{j=1;\text{While}[j\leq \text{fsize},\text{gradq1}[[j]]=\text{feat}[[j]];j=j+1];}\\
\pmb{j=1;\text{While}[j\leq \text{hlsize},\text{gradq1}[[\text{fsize}+j]]=\text{hlayer1}[[j]];j=j+1];}\\
\pmb{j=1;\text{While}[j\leq \text{fsize},\text{gradq2}[[j]]=\text{feat}[[j]];j=j+1];}\\
\pmb{j=1;\text{While}[j\leq \text{hlsize},\text{gradq2}[[\text{fsize}+j]]=\text{hlayer2}[[j]];j=j+1];}\\
\pmb{}\\
\pmb{\text{(* Calculate wth in the current state. *)}}\\
\pmb{\text{wth}=\text{mon}+(\text{cns}*\text{prices}[[i*5]]);}\\
\pmb{}\\
\pmb{\text{While}[(i+1)*5\leq \text{simlength},}\\
\pmb{}\\
\pmb{\text{(* Calculate Q1 and Q2 for every action in the given state. *)}}\\
\pmb{n=1;\text{While}[n\leq 19,}\\
\pmb{q=0;j=1;\text{While}[j\leq \text{hlsize}+\text{fsize},q=q+\text{wout1}[[n,j]]*\text{gradq1}[[j]];j=j+1];}\\
\pmb{\text{qarr1}[[n]]=q;}\\
\pmb{q=0;j=1;\text{While}[j\leq \text{hlsize}+\text{fsize},q=q+\text{wout2}[[n,j]]*\text{gradq2}[[j]];j=j+1];}\\
\pmb{\text{qarr2}[[n]]=q;}\\
\pmb{n=n+1];}\\
\pmb{}\\
\pmb{\text{(* Calculate the average of Q1 and Q2, which is needed to choose an action. *)}}\\
\pmb{\text{qarr}=(\text{qarr1}+\text{qarr2})/2.;}\\
\pmb{}\\
\pmb{\text{(* Update epsilon. *)}}\\
\pmb{\text{rand}=\text{RandomReal}[];}\\
\pmb{\text{If}[((\text{rand}\leq \text{probeps})\&\&(\text{ieps}\geq \text{Ceiling}[(\text{Exp}[1./0.2]-2.)/5.])),\text{ieps}=\text{Ceiling}[(\text{Exp}[1./0.2]-2.)/5.],\text{ieps}=\text{ieps}+1];}\\
\pmb{\text{eps}=1./\text{Log}[\text{ieps}*5.+2.];}\\
\pmb{}\\
\pmb{\text{(* Choose an action. *)}}\\
\pmb{\text{rand}=\text{RandomReal}[];}\\
\pmb{\text{If}[\text{rand}\leq \text{eps},a=\text{RandomInteger}[\{1,19\}],a=\text{Ordering}[\text{qarr},-1][[1]]];}\\
\pmb{}\\
\pmb{\text{(* Update alpha. *)}}\\
\pmb{\text{ialpha}=\text{ialpha}+1;}\\
\pmb{\text{alpha}=0.001+(1./2.)*(1.-0.001)*(1.+\text{Cos}[\text{Pi}*\text{ialpha}/1000.]);}\\
\pmb{}\\
\pmb{\text{(* Take the chosen action. *)}}\\
\pmb{\text{fff}=1.;\text{(* This tracks for insufficient mon or cns in order to reflect it in the reward later. *)}}\\
\pmb{\text{If}[a==1,\text{If}[\text{mon}<10.,\text{fff}=-1.;\text{Goto}[\text{fail}]];\text{cns}=\text{cns}+((99.9/100.)*(10./\text{prices}[[i*5]]));\text{mon}=\text{mon}-10.];}\\
\pmb{\text{If}[a==2,\text{If}[\text{mon}<20.,\text{fff}=-1.;\text{Goto}[\text{fail}]];\text{cns}=\text{cns}+((99.9/100.)*(20./\text{prices}[[i*5]]));\text{mon}=\text{mon}-20.];}\\
\pmb{\text{If}[a==3,\text{If}[\text{mon}<30.,\text{fff}=-1.;\text{Goto}[\text{fail}]];\text{cns}=\text{cns}+((99.9/100.)*(30./\text{prices}[[i*5]]));\text{mon}=\text{mon}-30.];}\\
\pmb{\text{If}[a==4,\text{If}[\text{mon}<40.,\text{fff}=-1.;\text{Goto}[\text{fail}]];\text{cns}=\text{cns}+((99.9/100.)*(40./\text{prices}[[i*5]]));\text{mon}=\text{mon}-40.];}\\
\pmb{\text{If}[a==5,\text{If}[\text{mon}<50.,\text{fff}=-1.;\text{Goto}[\text{fail}]];\text{cns}=\text{cns}+((99.9/100.)*(50./\text{prices}[[i*5]]));\text{mon}=\text{mon}-50.];}\\
\pmb{\text{If}[a==6,\text{If}[\text{mon}<60.,\text{fff}=-1.;\text{Goto}[\text{fail}]];\text{cns}=\text{cns}+((99.9/100.)*(60./\text{prices}[[i*5]]));\text{mon}=\text{mon}-60.];}\\
\pmb{\text{If}[a==7,\text{If}[\text{mon}<70.,\text{fff}=-1.;\text{Goto}[\text{fail}]];\text{cns}=\text{cns}+((99.9/100.)*(70./\text{prices}[[i*5]]));\text{mon}=\text{mon}-70.];}\\
\pmb{\text{If}[a==8,\text{If}[\text{mon}<80.,\text{fff}=-1.;\text{Goto}[\text{fail}]];\text{cns}=\text{cns}+((99.9/100.)*(80./\text{prices}[[i*5]]));\text{mon}=\text{mon}-80.];}\\
\pmb{\text{If}[a==9,\text{If}[\text{mon}<90.,\text{fff}=-1.;\text{Goto}[\text{fail}]];\text{cns}=\text{cns}+((99.9/100.)*(90./\text{prices}[[i*5]]));\text{mon}=\text{mon}-90.];}\\
\pmb{\text{If}[a==10,\text{If}[\text{prices}[[i*5]]*\text{cns}<10.,\text{fff}=-1.;\text{Goto}[\text{fail}]];\text{mon}=\text{mon}+((99.9/100.)*(10.));\text{cns}=\text{cns}-(10./\text{prices}[[i*5]])];}\\
\pmb{\text{If}[a==11,\text{If}[\text{prices}[[i*5]]*\text{cns}<20.,\text{fff}=-1.;\text{Goto}[\text{fail}]];\text{mon}=\text{mon}+((99.9/100.)*(20.));\text{cns}=\text{cns}-(20./\text{prices}[[i*5]])];}\\
\pmb{\text{If}[a==12,\text{If}[\text{prices}[[i*5]]*\text{cns}<30.,\text{fff}=-1.;\text{Goto}[\text{fail}]];\text{mon}=\text{mon}+((99.9/100.)*(30.));\text{cns}=\text{cns}-(30./\text{prices}[[i*5]])];}\\
\pmb{\text{If}[a==13,\text{If}[\text{prices}[[i*5]]*\text{cns}<40.,\text{fff}=-1.;\text{Goto}[\text{fail}]];\text{mon}=\text{mon}+((99.9/100.)*(40.));\text{cns}=\text{cns}-(40./\text{prices}[[i*5]])];}\\
\pmb{\text{If}[a==14,\text{If}[\text{prices}[[i*5]]*\text{cns}<50.,\text{fff}=-1.;\text{Goto}[\text{fail}]];\text{mon}=\text{mon}+((99.9/100.)*(50.));\text{cns}=\text{cns}-(50./\text{prices}[[i*5]])];}\\
\pmb{\text{If}[a==15,\text{If}[\text{prices}[[i*5]]*\text{cns}<60.,\text{fff}=-1.;\text{Goto}[\text{fail}]];\text{mon}=\text{mon}+((99.9/100.)*(60.));\text{cns}=\text{cns}-(60./\text{prices}[[i*5]])];}\\
\pmb{\text{If}[a==16,\text{If}[\text{prices}[[i*5]]*\text{cns}<70.,\text{fff}=-1.;\text{Goto}[\text{fail}]];\text{mon}=\text{mon}+((99.9/100.)*(70.));\text{cns}=\text{cns}-(70./\text{prices}[[i*5]])];}\\
\pmb{\text{If}[a==17,\text{If}[\text{prices}[[i*5]]*\text{cns}<80.,\text{fff}=-1.;\text{Goto}[\text{fail}]];\text{mon}=\text{mon}+((99.9/100.)*(80.));\text{cns}=\text{cns}-(80./\text{prices}[[i*5]])];}\\
\pmb{\text{If}[a==18,\text{If}[\text{prices}[[i*5]]*\text{cns}<90.,\text{fff}=-1.;\text{Goto}[\text{fail}]];\text{mon}=\text{mon}+((99.9/100.)*(90.));\text{cns}=\text{cns}-(90./\text{prices}[[i*5]])];}\\
\pmb{\text{(* } \text{If}[a==19,\text{HOLD}]; \text{ *)}}\\
\pmb{\text{Label}[\text{fail}];}\\
\pmb{\text{If}[((\text{mon}<0.)\|(\text{cns}<0.)),\text{Print}[\text{{``}Negative mon or cns!{''}}]];\text{(* Just to be sure. *)}}\\
\pmb{}\\
\pmb{\text{(* Observe the reward (by calculating wth in the next state). *)}}\\
\pmb{\text{wthnew}=\text{mon}+(\text{cns}*\text{prices}[[(i+1)*5]]);}\\
\pmb{\text{rew}=\text{wthnew}-\text{wth};}\\
\pmb{\text{If}[\text{fff}<0.,}\\
\pmb{\text{rew}=\text{rew}-((\text{rew}/2.){}^{\wedge}2)-0.1,}\\
\pmb{\text{rew}=\text{rew}-((\text{rew}/2.){}^{\wedge}2)];}\\
\pmb{}\\
\pmb{\text{(* Observe the prices in the next state. *)}}\\
\pmb{j=1;\text{While}[j\leq \text{vsize},\text{pr}[[j]]=\text{prices}[[(i+1)*5-\text{vsize}+j]];j=j+1];}\\
\pmb{}\\
\pmb{\text{(* Calculate RSI for 15 points. *)}}\\
\pmb{j=1;\text{While}[j\leq 10,\text{rsip}[[j]]=\text{rsip}[[j+5]];j=j+1];}\\
\pmb{j=1;\text{While}[j\leq \text{vsize},\text{rsip}[[10+j]]=\text{pr}[[j]];j=j+1];}\\
\pmb{j=1;\text{While}[j\leq 14,}\\
\pmb{\text{If}[\text{rsip}[[j+1]]>\text{rsip}[[j]],\text{rsiu}[[j]]=\text{rsip}[[j+1]]-\text{rsip}[[j]];\text{rsid}[[j]]=0.];}\\
\pmb{\text{If}[\text{rsip}[[j+1]]<\text{rsip}[[j]],\text{rsid}[[j]]=\text{rsip}[[j]]-\text{rsip}[[j+1]];\text{rsiu}[[j]]=0.];}\\
\pmb{\text{If}[\text{rsip}[[j+1]]==\text{rsip}[[j]],\text{rsiu}[[j]]=0.;\text{rsid}[[j]]=0.];}\\
\pmb{j=j+1];}\\
\pmb{\text{If}[\text{Mean}[\text{rsid}]==0.,\text{rsi}=100.,\text{rsi}=100.-(100./(1.+(\text{Mean}[\text{rsiu}]/\text{Mean}[\text{rsid}])))];}\\
\pmb{}\\
\pmb{\text{(* Check for the three possible terminal states. *)}}\\
\pmb{\text{If}[\text{mon}>\text{mlim},\text{(* Terminal state 1. *)}}\\
\pmb{\text{rew}=\text{rew}+((\text{mon}-\text{mlim})*0.34);}\\
\pmb{\text{rand}=\text{RandomReal}[];}\\
\pmb{\text{If}[\text{rand}\leq 0.5,}\\
\pmb{\text{wout1}[[a]]=\text{wout1}[[a]]+\text{alpha}*(\text{rew}-\text{qarr1}[[a]])*\text{gradq1};}\\
\pmb{\text{max1}=\text{Max}[\text{max1},\text{Norm}[\text{wout1}[[a]]]];}\\
\pmb{\text{If}[\text{Norm}[\text{wout1}[[a]]]>1.,\text{wout1}[[a]]=\text{wout1}[[a]]/\text{max1}],}\\
\pmb{\text{wout2}[[a]]=\text{wout2}[[a]]+\text{alpha}*(\text{rew}-\text{qarr2}[[a]])*\text{gradq2};}\\
\pmb{\text{max2}=\text{Max}[\text{max2},\text{Norm}[\text{wout2}[[a]]]];}\\
\pmb{\text{If}[\text{Norm}[\text{wout2}[[a]]]>1.,\text{wout2}[[a]]=\text{wout2}[[a]]/\text{max2}]];}\\
\pmb{\text{mdf}=\text{mon}-\text{mlim};}\\
\pmb{\text{sav}=\text{sav}+(\text{mdf}*0.34);}\\
\pmb{\text{res}=\text{res}+(\text{mdf}*0.33);}\\
\pmb{\text{mon}=\text{mlim}+(\text{mdf}*0.33);}\\
\pmb{\text{mlim}=\text{mon}+\text{mdf};}\\
\pmb{i=i+2;}\\
\pmb{\text{If}[i*5>\text{simlength},\text{Goto}[\text{break}]];}\\
\pmb{\text{Goto}[\text{episode}],}\\
\pmb{\text{If}[((\text{wthnew}<\text{mlimn})\&\&(\text{qarr}[[a]]>0.)\&\&(\text{rsi}>70.)),\text{(* Terminal state 2. *)}}\\
\pmb{\text{rand}=\text{RandomReal}[];}\\
\pmb{\text{If}[\text{rand}\leq 0.5,}\\
\pmb{\text{wout1}[[a]]=\text{wout1}[[a]]+\text{alpha}*(\text{rew}-\text{qarr1}[[a]])*\text{gradq1};}\\
\pmb{\text{max1}=\text{Max}[\text{max1},\text{Norm}[\text{wout1}[[a]]]];}\\
\pmb{\text{If}[\text{Norm}[\text{wout1}[[a]]]>1.,\text{wout1}[[a]]=\text{wout1}[[a]]/\text{max1}],}\\
\pmb{\text{wout2}[[a]]=\text{wout2}[[a]]+\text{alpha}*(\text{rew}-\text{qarr2}[[a]])*\text{gradq2};}\\
\pmb{\text{max2}=\text{Max}[\text{max2},\text{Norm}[\text{wout2}[[a]]]];}\\
\pmb{\text{If}[\text{Norm}[\text{wout2}[[a]]]>1.,\text{wout2}[[a]]=\text{wout2}[[a]]/\text{max2}]];}\\
\pmb{\text{mon}=\text{mon}+(\text{res}/2.);\text{res}=\text{res}-(\text{res}/2.);}\\
\pmb{\text{If}[\text{mon}\geq \text{mlimn},\text{mlim}=\text{mon},\text{mlim}=\text{mlimn}];}\\
\pmb{i=i+2;}\\
\pmb{\text{If}[i*5>\text{simlength},\text{Goto}[\text{break}]];}\\
\pmb{\text{Goto}[\text{episode}]];}\\
\pmb{\text{If}[((\text{wthnew}\geq \text{mlimn})\&\&(\text{qarr}[[a]]<0.)\&\&(\text{rsi}<30.)),\text{(* Terminal state 3. *)}}\\
\pmb{\text{rand}=\text{RandomReal}[];}\\
\pmb{\text{If}[\text{rand}\leq 0.5,}\\
\pmb{\text{wout1}[[a]]=\text{wout1}[[a]]+\text{alpha}*(\text{rew}-\text{qarr1}[[a]])*\text{gradq1};}\\
\pmb{\text{max1}=\text{Max}[\text{max1},\text{Norm}[\text{wout1}[[a]]]];}\\
\pmb{\text{If}[\text{Norm}[\text{wout1}[[a]]]>1.,\text{wout1}[[a]]=\text{wout1}[[a]]/\text{max1}],}\\
\pmb{\text{wout2}[[a]]=\text{wout2}[[a]]+\text{alpha}*(\text{rew}-\text{qarr2}[[a]])*\text{gradq2};}\\
\pmb{\text{max2}=\text{Max}[\text{max2},\text{Norm}[\text{wout2}[[a]]]];}\\
\pmb{\text{If}[\text{Norm}[\text{wout2}[[a]]]>1.,\text{wout2}[[a]]=\text{wout2}[[a]]/\text{max2}]];}\\
\pmb{\text{mlim}=\text{wthnew};}\\
\pmb{i=i+2;}\\
\pmb{\text{If}[i*5>\text{simlength},\text{Goto}[\text{break}]];}\\
\pmb{\text{Goto}[\text{episode}]]];}\\
\pmb{}\\
\pmb{\text{(* Calculate the average volumes. *)}}\\
\pmb{\text{cav}=(\text{volumes}[[(i+1)*5-4]]+\text{volumes}[[(i+1)*5-3]]+\text{volumes}[[(i+1)*5-2]]+\text{volumes}[[(i+1)*5-1]]+\text{volumes}[[(i+1)*5]])/5.;}\\
\pmb{j=1;\text{While}[j\leq 19,\text{mv}[[j]]=\text{mv}[[j+1]];j=j+1];}\\
\pmb{\text{mv}[[j]]=\text{cav};}\\
\pmb{\text{av}=\text{Mean}[\text{mv}];}\\
\pmb{}\\
\pmb{\text{(* Relative price movements. *)}}\\
\pmb{j=1;\text{While}[j\leq \text{vsize}-1,}\\
\pmb{\text{nmd1}[[j]]=(\text{pr}[[j+1]]-\text{pr}[[j]])/\text{pr}[[j]];}\\
\pmb{j=j+1];}\\
\pmb{j=1;\text{While}[j\leq \text{vsize}-2,}\\
\pmb{\text{nmd2}[[j]]=(\text{nmd1}[[j+1]]-\text{nmd1}[[j]])/\text{Abs}[\text{nmd1}[[j]]];}\\
\pmb{j=j+1];}\\
\pmb{j=1;\text{While}[j\leq \text{vsize}-3,}\\
\pmb{\text{nmd3}[[j]]=(\text{nmd2}[[j+1]]-\text{nmd2}[[j]])/\text{Abs}[\text{nmd2}[[j]]];}\\
\pmb{j=j+1];}\\
\pmb{j=1;\text{While}[j\leq \text{vsize}-4,}\\
\pmb{\text{nmd4}[[j]]=(\text{nmd3}[[j+1]]-\text{nmd3}[[j]])/\text{Abs}[\text{nmd3}[[j]]];}\\
\pmb{j=j+1];}\\
\pmb{}\\
\pmb{\text{(* Construct the feature vector (same for all actions). *)}}\\
\pmb{\text{feat}[[1]]=0.;}\\
\pmb{j=1;k=1;\text{While}[j\leq \text{vsize},\text{feat}[[k+1]]=\text{pr}[[j]];k=k+1;j=j+1];}\\
\pmb{\text{feat}[[k+1]]=\text{ipr};k=k+1;}\\
\pmb{\text{feat}[[k+1]]=(\text{pr}[[\text{vsize}]]-\text{ipr})/\text{ipr};k=k+1;}\\
\pmb{\text{feat}[[k+1]]=\text{mon};k=k+1;}\\
\pmb{\text{feat}[[k+1]]=\text{cns};k=k+1;}\\
\pmb{\text{feat}[[k+1]]=\text{cav};k=k+1;}\\
\pmb{\text{feat}[[k+1]]=\text{av};k=k+1;}\\
\pmb{\text{feat}[[k+1]]=(\text{cav}-\text{av})/\text{av};k=k+1;}\\
\pmb{\text{feat}[[k+1]]=(\text{volumes}[[(i+1)*5]]-\text{av})/\text{av};k=k+1;}\\
\pmb{\text{feat}[[k+1]]=(\text{volumes}[[(i+1)*5]]-\text{cav})/\text{cav};k=k+1;}\\
\pmb{\text{feat}[[k+1]]=\text{rsi};k=k+1;}\\
\pmb{j=1;\text{While}[j\leq \text{vsize}-1,\text{feat}[[k+1]]=\text{nmd1}[[j]];k=k+1;j=j+1];}\\
\pmb{j=1;\text{While}[j\leq \text{vsize}-2,\text{feat}[[k+1]]=\text{nmd2}[[j]];k=k+1;j=j+1];}\\
\pmb{j=1;\text{While}[j\leq \text{vsize}-3,\text{feat}[[k+1]]=\text{nmd3}[[j]];k=k+1;j=j+1];}\\
\pmb{j=1;\text{While}[j\leq \text{vsize}-4,\text{feat}[[k+1]]=\text{nmd4}[[j]];k=k+1;j=j+1];}\\
\pmb{\text{feat}[[k+1]]=\text{mlim};k=k+1;}\\
\pmb{}\\
\pmb{\text{(* Rescale the feature vector, so its norm is 6. *)}}\\
\pmb{\text{feat}=6.*(\text{feat}/\text{Norm}[\text{feat}]);}\\
\pmb{\text{feat}[[1]]=1.;}\\
\pmb{}\\
\pmb{\text{(* Output of the hidden layers. *)}}\\
\pmb{n=1;\text{While}[n\leq \text{hlsize},}\\
\pmb{q=0;j=1;\text{While}[j\leq \text{fsize},q=q+\text{whi1}[[n,j]]*\text{feat}[[j]];j=j+1];}\\
\pmb{\text{hlayer1}[[n]]=1./(1.+\text{Exp}[-q]);}\\
\pmb{q=0;j=1;\text{While}[j\leq \text{fsize},q=q+\text{whi2}[[n,j]]*\text{feat}[[j]];j=j+1];}\\
\pmb{\text{hlayer2}[[n]]=1./(1.+\text{Exp}[-q]);}\\
\pmb{n=n+1];}\\
\pmb{}\\
\pmb{\text{(* Gradients of the Q-functions, taken in the next state (same for all actions), which are just}}\\
\pmb{\text{concatenations of the feature vector with the hidden layer outputs. *)}}\\
\pmb{j=1;\text{While}[j\leq \text{fsize},\text{gradq1new}[[j]]=\text{feat}[[j]];j=j+1];}\\
\pmb{j=1;\text{While}[j\leq \text{hlsize},\text{gradq1new}[[\text{fsize}+j]]=\text{hlayer1}[[j]];j=j+1];}\\
\pmb{j=1;\text{While}[j\leq \text{fsize},\text{gradq2new}[[j]]=\text{feat}[[j]];j=j+1];}\\
\pmb{j=1;\text{While}[j\leq \text{hlsize},\text{gradq2new}[[\text{fsize}+j]]=\text{hlayer2}[[j]];j=j+1];}\\
\pmb{}\\
\pmb{\text{(* Calculate Q1 and Q2 for every action in the next state. *)}}\\
\pmb{n=1;\text{While}[n\leq 19,}\\
\pmb{q=0;j=1;\text{While}[j\leq \text{hlsize}+\text{fsize},q=q+\text{wout1}[[n,j]]*\text{gradq1new}[[j]];j=j+1];}\\
\pmb{\text{qarr1new}[[n]]=q;}\\
\pmb{q=0;j=1;\text{While}[j\leq \text{hlsize}+\text{fsize},q=q+\text{wout2}[[n,j]]*\text{gradq2new}[[j]];j=j+1];}\\
\pmb{\text{qarr2new}[[n]]=q;}\\
\pmb{n=n+1];}\\
\pmb{}\\
\pmb{\text{(* Update the output weights. *)}}\\
\pmb{\text{rand}=\text{RandomReal}[];}\\
\pmb{\text{If}[\text{rand}\leq 0.5,}\\
\pmb{\text{amax}=\text{Ordering}[\text{qarr1new},-1][[1]];}\\
\pmb{\text{wout1}[[a]]=\text{wout1}[[a]]+\text{alpha}*(\text{rew}+\text{gamma}*\text{qarr2new}[[\text{amax}]]-\text{qarr1}[[a]])*\text{gradq1};}\\
\pmb{\text{max1}=\text{Max}[\text{max1},\text{Norm}[\text{wout1}[[a]]]];}\\
\pmb{\text{If}[\text{Norm}[\text{wout1}[[a]]]>1.,\text{wout1}[[a]]=\text{wout1}[[a]]/\text{max1}],}\\
\pmb{\text{amax}=\text{Ordering}[\text{qarr2new},-1][[1]];}\\
\pmb{\text{wout2}[[a]]=\text{wout2}[[a]]+\text{alpha}*(\text{rew}+\text{gamma}*\text{qarr1new}[[\text{amax}]]-\text{qarr2}[[a]])*\text{gradq2};}\\
\pmb{\text{max2}=\text{Max}[\text{max2},\text{Norm}[\text{wout2}[[a]]]];}\\
\pmb{\text{If}[\text{Norm}[\text{wout2}[[a]]]>1.,\text{wout2}[[a]]=\text{wout2}[[a]]/\text{max2}]];}\\
\pmb{}\\
\pmb{\text{(* Make the next value of wth and the next states current. *)}}\\
\pmb{\text{wth}=\text{wthnew};}\\
\pmb{\text{gradq1}=\text{gradq1new};}\\
\pmb{\text{gradq2}=\text{gradq2new};}\\
\pmb{}\\
\pmb{i=i+1;}\\
\pmb{}\\
\pmb{];}\\
\pmb{}\\
\pmb{\text{Label}[\text{break}];\text{(* Label used to exit the loop, if there are no more points in the test data. *)}}\\
\pmb{\text{savarray}[[m]]=\text{sav};\text{(* Write sav at the end of the current run. *)}}\\
\pmb{\text{twtharray}[[m]]=\text{sav}+\text{wthnew}+\text{res};\text{(* Write twth at the end of the current run. *)}}\\
\pmb{}\\
\pmb{m=m+1}\\
\pmb{}\\
\pmb{];}\\
\pmb{}\\
\pmb{\text{Export}[\text{{``}sav.txt{''}},\text{savarray}]}\\
\pmb{\text{Export}[\text{{``}total.txt{''}},\text{twtharray}]}\)}

\end{appendices}

\end{document}